

\magnification=\magstep1 \openup1\jot
\font\eightrm=cmr8
\def\boxit#1{\vbox{\hrule\hbox{\vrule\kern3pt\vbox{\kern3pt#1\kern3pt}
\kern3pt\vrule}\hrule}}
\input epsf.tex

\centerline{\bf 1. Classical Theory} \smallskip \centerline{S. W. Hawking}
\bigskip

In these lectures Roger Penrose and I will put forward our related but rather
different viewpoints on the nature of space and time. We shall speak
alternately and shall give three lectures each, followed by a discussion on our
different approaches. I should emphasize that these will be technical lectures.
We shall assume a basic knowledge of general relativity and quantum theory.

There is a short article by Richard Feynman describing his experiences at a
conference on general relativity. I think it was the Warsaw conference in
1962. It commented very unfavorably on the general competence of the people
there and the relevance of what they were doing. That general relativity soon
acquired a much better reputation, and more interest, is in a considerable
measure because of Roger's work.  Up to then, general relativity had been
formulated as a messy set of partial differential equations in a single
coordinate system. People were so pleased when they found a solution that they
didn't care that it probably had no physical significance. However, Roger
brought in modern concepts like spinors and global methods. He was the first
to show that one could discover general properties without solving the
equations exactly. It was his first singularity theorem that introduced me to
the study of causal structure and inspired my classical work on singularities
and black holes.

I think Roger and I pretty much agree on the classical work. However, we differ
in our approach to quantum gravity and indeed to quantum theory itself.
Although I'm regarded as a dangerous radical by particle physicists for
proposing that there may be loss of quantum coherence I'm definitely a
conservative compared to Roger. I take the positivist viewpoint that a physical
theory is just a mathematical model and that it is meaningless to ask whether
it corresponds to reality. All that one can ask is that its predictions should
be in agreement with observation. I think Roger is a Platonist at heart but he
must answer for himself.

Although there have been suggestions that spacetime may have a discrete
structure I see no reason to abandon the continuum theories that have been so
successful. General relativity is a beautiful theory that agrees with every
observation that has been made. It may require modifications on the Planck
scale but I don't think that will affect many of the predictions that can be
obtained from it. It may be only a low energy approximation to some more
fundemental theory, like string theory, but I think string theory has been over
sold. First of all, it is not clear that general relativity, when combined with
various other fields in a supergravity theory, can not give a sensible quantum
theory. Reports of the death of supergravity are exaggerations. One year
everyone believed that supergravity was finite. The next year the fashion
changed and everyone said that supergravity was bound to have divergences even
though none had actually been found. My second reason for not discussing string
theory is that it has not made any testable predictions. By contrast, the
straight forward application of quantum theory to general relativity, which I
will be talking about, has already made two testable predictions. One of these
predictions, the development of small perturbations during inflation, seems to
be confirmed by recent observations of fluctuations in the microwave
background.
The other prediction, that black holes should radiate thermally, is testable in
principle. All we have to do is find a primordial black hole. Unfortunately,
there don't seem many around in this neck of the woods. If there had been we
would know how to quantize gravity.

Neither of these predictions will be changed even if string theory is the
ultimate theory of nature. But string theory, at least at its current state of
development, is quite incapable of making these predictions except by appealing
to general relativity as the low energy effective theory. I suspect this may
always be the case and that there may not be any observable predictions of
string theory that can not also be predicted from general relativity or
supergravity.  If this is true it raises the question of whether string theory
is a genuine scientific theory. Is mathematical beauty and completeness enough
in the absence of distinctive observationally tested predictions. Not that
string theory in its present form is either beautiful or complete.

For these reasons, I shall talk about general relativity in these lectures. I
shall concentrate on two areas where gravity seems to lead to features that
are completely different from other field theories. The first is the idea that
gravity should cause spacetime to have a begining and maybe an end. The second
is the discovery that there seems to be intrinsic gravitational entropy that
is not the result of coarse graining. Some people have claimed that these
predictions are just artifacts of the semi classical approximation. They say
that string theory, the true quantum theory of gravity, will smear out the
singularities and will introduce correlations in the radiation from black
holes so that it is only approximately thermal in the coarse grained sense. It
would be rather boring if this were the case.  Gravity would be just like any
other field. But I believe it is distinctively different, because it shapes the
arena in which it acts, unlike other fields which act in a fixed spacetime
background. It is this that leads to the possibility of time having a begining.
It also leads to  regions of the universe which one can't observe, which in
turn
gives rise to the concept of gravitational entropy as a measure of what we
can't know.

In this lecture I shall review the work in classical general relativity that
leads to these ideas. In the second and third lectures I shall show how they
are changed and extended when one goes to quantum theory. Lecture two will be
about black holes and lecture three will be on quantum cosmology.

The crucial technique for investigating singularities and black holes that was
introduced by Roger, and which I helped develop, was the study of the global
causal structure of spacetime.

\midinsert
\hskip 0.8in
\epsfbox{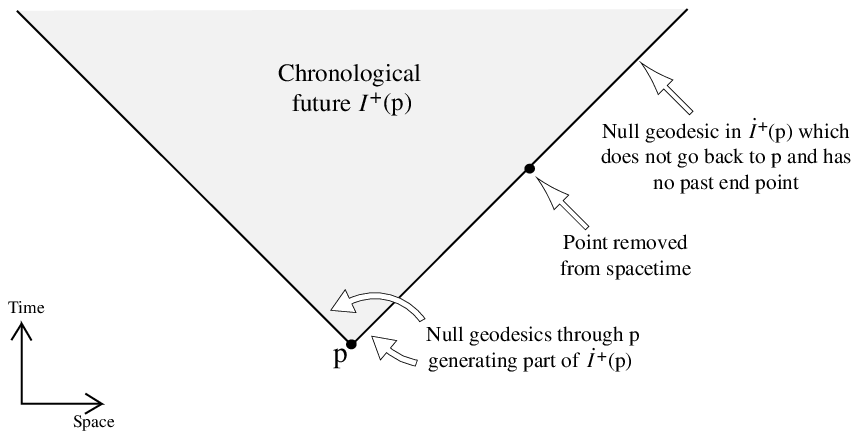}
\endinsert

\noindent Define $I^+(p)$ to be the set
of all points of the spacetime $M$ that can be reached from $p$ by future
directed time like curves. One can think of $I^+(p)$ as the set of all events
that can be influenced by what happens at $p$.  There are similar definitions
in
which plus is replaced by minus and future by past. I shall regard such
definitions as self evident.

\midinsert
\hskip 0.1in
\epsfbox{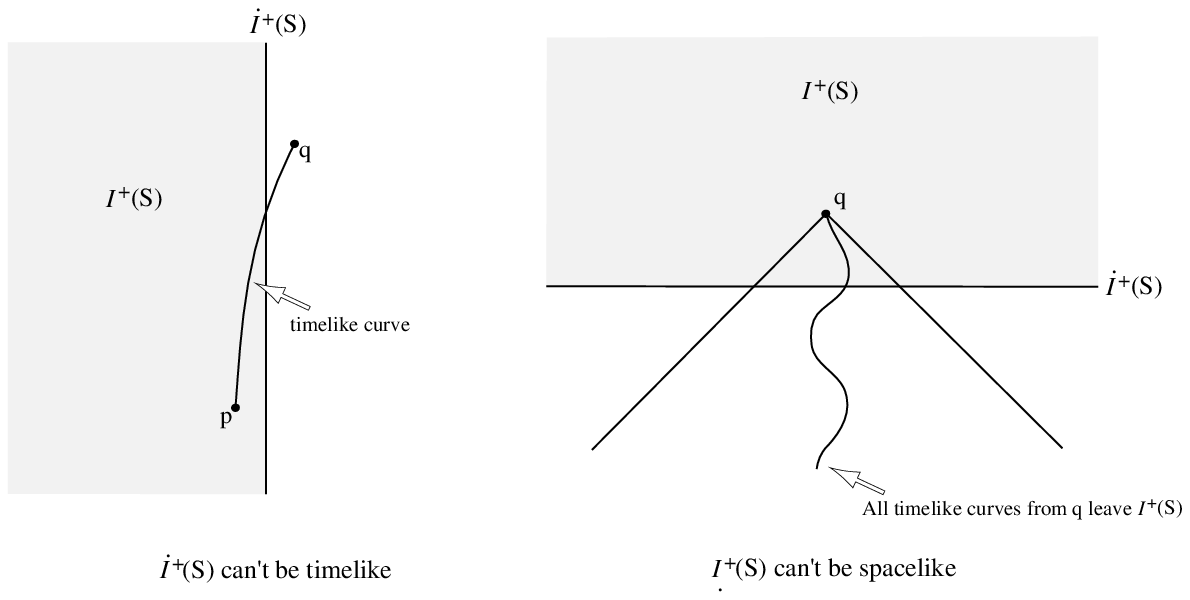}
\endinsert

One now considers the boundary ${\dot I}^+(S)$ of the future of a set $S$.  It
is fairly easy to see that this boundary can not be time like. For in that
case,
a point $q$ just outside the boundary would be to the future of a point $p$
just inside. Nor can the boundary of the future be space like, except at the
set $S$ itself. For in that case every past directed curve from a point $q$,
just
to the future of the boundary, would cross the boundary and leave the future of
$S$. That would be a contradiction with the fact that $q$ is in the future of
$S$.

\midinsert
\hskip 0.8in
\epsfbox{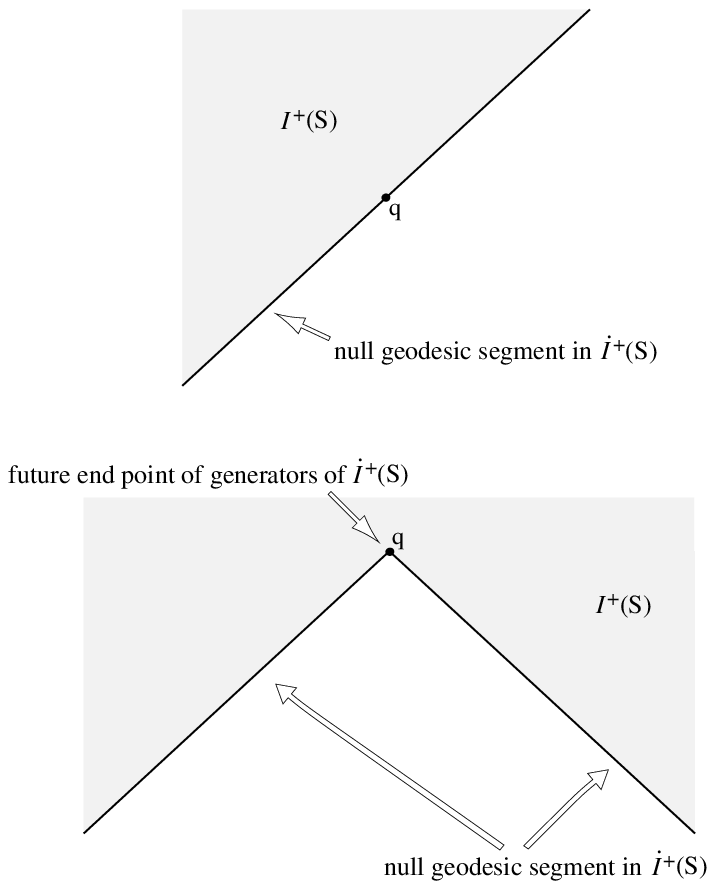}
\endinsert

One therefore concludes that the boundary of the future is null apart from at
$S$ itself. More precisely, if $q$ is in the boundary of the future but is not
in
the closure of $S$ there is a past directed null geodesic segment through $q$
lying in the boundary. There may be more than one null geodesic segment through
$q$ lying in the boundary, but in that case $q$ will be a future end point of
the
segments. In other words, the boundary of the future of $S$ is generated by
null
geodesics that have a future end point in the boundary and pass into the
interior of the future if they intersect another generator. On the other hand,
the null geodesic generators can have past end points only on $S$. It is
possible, however, to have spacetimes in which there are generators of the
boundary of the future of a set $S$ that never intersect $S$. Such generators
can
have no past end point.

\midinsert
\hskip 0.5in
\epsfbox{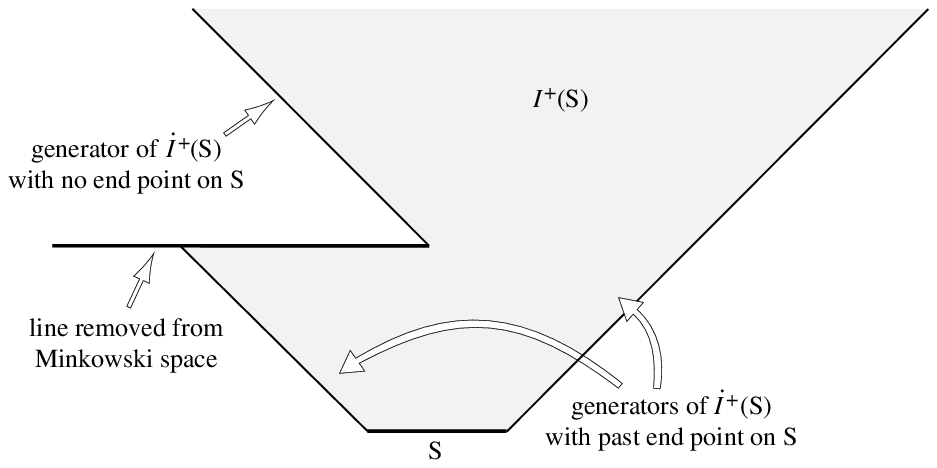}
\endinsert

A simple example of this is Minkowski space with a
horizontal line segment removed. If the set $S$ lies to the past of the
horizontal
line, the line will cast a shadow and there will be points just to the future
of the line that are not in the future of $S$. There will be a generator of the
boundary of the future of $S$ that goes back to the end of the horizontal line.
However, as the end point of the horizontal line has been removed from
spacetime, this generator of the boundary will
have no past end point. This
spacetime is incomplete, but one can cure this by multiplying the metric by a
suitable conformal factor near the end of the horizontal line. Although spaces
like this are very artificial they are important in showing how careful you
have to be in the study of causal structure. In fact Roger Penrose, who was one
of my PhD examiners, pointed out that a space like that I have just described
was a counter example to some of the claims I made in my thesis.

To show that each generator of the boundary of the future has a past end point
on the set one has to impose some global condition on the causal structure. The
strongest and physically most important condition is that of global
hyperbolicity.

\midinsert
\hskip 0.8in
\epsfbox{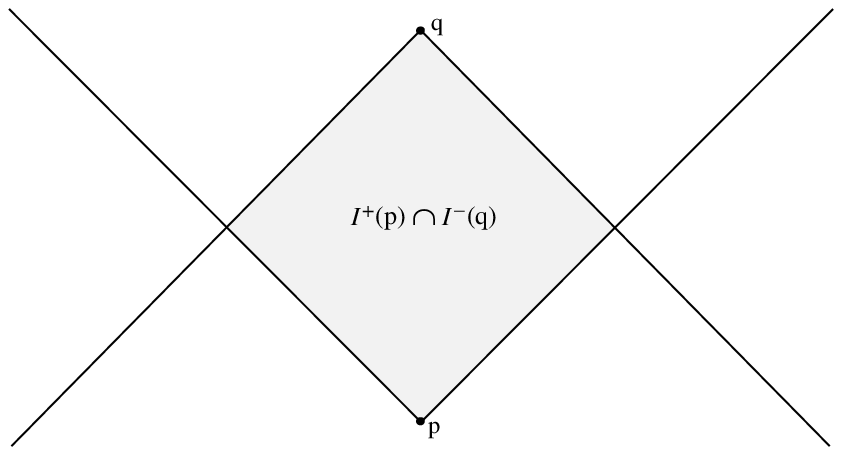}
\endinsert

\noindent An open set $U$ is said to be globally hyperbolic if:
\item{1)} for every pair of points $p$ and $q$ in $U$ the intersection of the
future of $p$ and
the past of $q$ has compact closure. In other words, it is a bounded diamond
shaped region.
\item{2)} strong causality holds on $U$. That is there are no
closed or almost closed time like curves contained in $U$.

\midinsert
\hskip 1.0in
\epsfbox{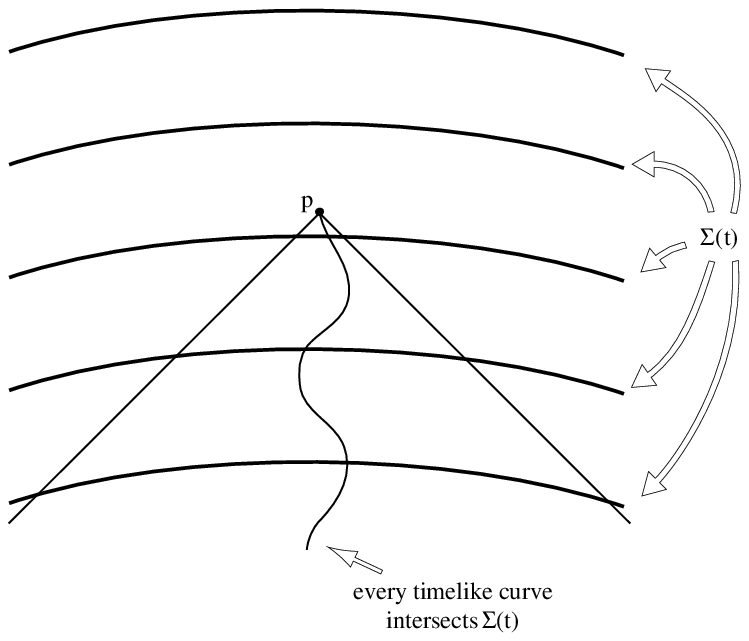}
\endinsert

The physical significance of global hyperbolicity comes from the fact that it
implies that there is a family of Cauchy surfaces $\Sigma (t)$ for $U$. A
Cauchy surface for $U$ is a space like or null surface that intersects every
time like curve in $U$ once and once only. One can predict what will happen in
$U$
from data on the Cauchy surface, and one can formulate a well behaved quantum
field theory on a globally hyperbolic background. Whether one can formulate a
sensible quantum field theory on a non globally hyperbolic background is less
clear. So global hyperbolicity may be a physical necessity. But my view point
is that one shouldn't assume it because that may be ruling out something that
gravity is trying to tell us. Rather one should deduce that certain regions of
spacetime are globally hyperbolic from other physically reasonable assumptions.

The significance of global hyperbolicity for singularity theorems stems from
the following.

\midinsert
\hskip 0.8in
\epsfbox{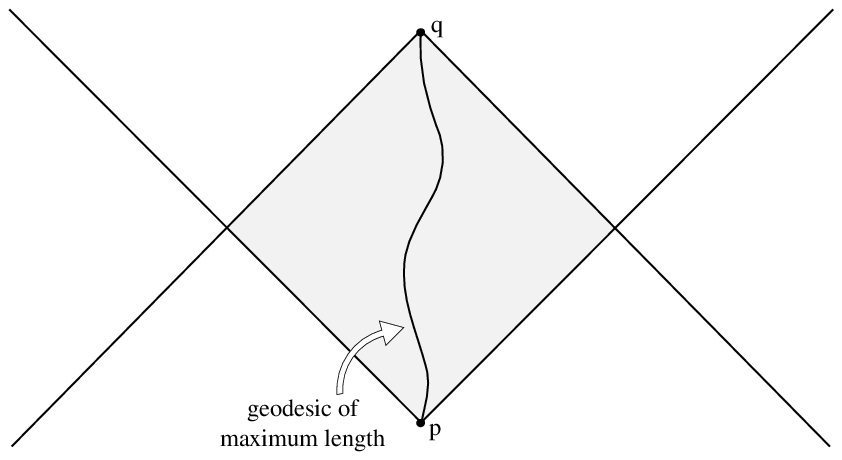}
\endinsert

\noindent Let $U$ be globally hyperbolic and let $p$ and $q$ be points of $U$
that can be joined by a time like or null curve. Then there is a time like or
null geodesic between $p$ and $q$ which maximizes the length of time like or
null
curves from $p$ to $q$. The method of proof is to show the space of all time
like
or null curves from $p$ to $q$ is compact in a certain topology. One then shows
that the length of the curve is an upper semi continuous function on this
space. It must therefore attain its maximum and the curve of maximum length
will be a geodesic because otherwise a small variation will give a longer
curve.

\midinsert
\hskip 0.1in
\epsfbox{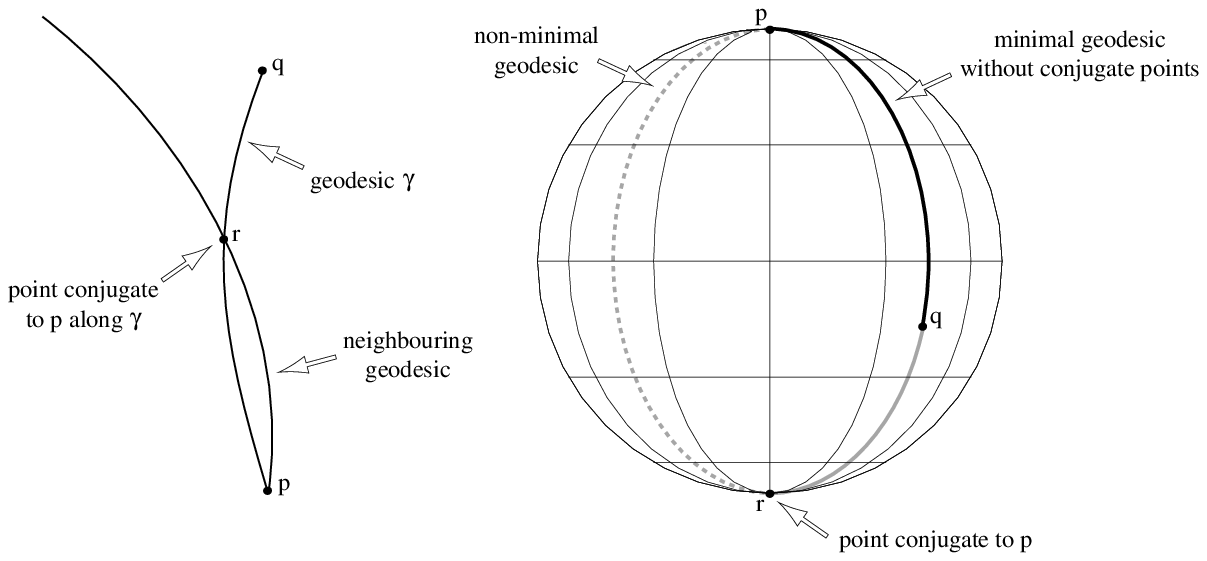}
\endinsert

One can now consider the second variation of the length of a geodesic $\gamma$.
One can show that $\gamma$ can be varied to a longer curve if there is an
infinitesimally neighbouring geodesic from $p$ which intersects $\gamma$ again
at a
point $r$ between $p$ and $q$. The point $r$ is said to be conjugate to $p$.
One can
illustrate this by considering two points $p$ and $q$ on the surface of the
Earth. Without loss of generality one can take $p$ to be at the north pole.
Because the Earth has a positive definite metric rather than a Lorentzian one,
there is a geodesic of minimal length, rather than a geodesic of maximum
length.
This minimal geodesic will be a line of longtitude running from the north pole
to the point $q$. But there will be another geodesic from $p$ to $q$ which runs
down
the back from the north pole to the south pole and then up to $q$. This
geodesic contains a point conjugate to $p$ at the south pole where all the
geodesics from $p$ intersect. Both geodesics from $p$ to $q$ are stationary
points of
the length under a small variation. But now in a positive definite metric the
second variation of a geodesic containing a conjugate point can give a shorter
curve from $p$ to $q$. Thus, in the example of the Earth, we can deduce that
the
geodesic that goes down to the south pole and then comes up is not the
shortest curve from $p$ to $q$. This example is very obvious. However, in the
case
of spacetime one can show that under certain assumptions there ought to be a
globally hyperbolic region in which there ought to be conjugate points on
every geodesic between two points.  This establishes a contradiction which
shows that the assumption of geodesic completeness, which can be taken as a
definition of a non singular spacetime, is false.

The reason one gets conjugate points in spacetime is that gravity is an
attractive force. It therefore curves spacetime in such a way that neighbouring
geodesics are bent towards each other rather than away. One can see this from
the Raychaudhuri or Newman-Penrose equation, which I will write in a unified
form.

\midinsert
\hsize=4.75in
\indent
\boxit{\smallskip
{\bf Raychaudhuri - Newman - Penrose equation}\smallskip
$${d \rho\over d v} ~~ = ~~ \rho^2 ~~ + ~~ \sigma^{ij}\sigma_{ij}
{}~~ + ~~{1\over n}R_{ab}l^a l^b$$
\itemitem{where} $n~=~2$ for null geodesics\smallskip
\itemitem{} $n~=~3$ for timelike geodesics\smallskip
\smallskip}
\hsize=5.4in
\endinsert

\noindent Here $v$ is an affine parameter along a congruence of geodesics, with
tangent vector $l^a$ which are hypersurface orthogonal. The quantity $\rho$ is
the
average rate of convergence of the geodesics, while $\sigma$ measures the
shear.
The term $R_{ab}l^al^b$ gives the direct gravitational effect of the matter
on the convergence of the geodesics.

\midinsert
\hsize=4.75in
\indent
\boxit{\smallskip
{\bf Einstein equation}
$$R_{ab} ~~ - ~~ {1\over 2}g_{ab} R ~~ = ~~ 8\pi T_{ab}$$
\medskip
{\bf Weak Energy Condition}
$$T_{ab}v^av^b ~~ \ge ~~ 0$$
\indent{for any timelike vector $v^a$.}
\smallskip}
\hsize=5.4in
\endinsert

\noindent By the Einstein equations, it will be non
negative for any null vector $l^a$ if the matter obeys the so called weak
energy
condition. This says that the energy density $T_{00}$ is non negative in any
frame.
The weak energy condition is obeyed by the classical energy momentum tensor of
any reasonable matter, such as a scalar or electro magnetic field or a fluid
with a reasonable equation of state. It may not however be satisfied locally
by the quantum mechanical expectation value of the energy momentum tensor. This
will be relevant in my second and third lectures.

Suppose the weak energy condition holds, and that the null geodesics from a
point $p$ begin to converge again and that $\rho$ has the positive value
$\rho_0$.
Then the Newman Penrose equation would imply that the convergence $\rho$ would
become infinite at a point $q$ within an affine parameter distance
${1\over\rho_0}$
if the null geodesic can be extended that far.

\midinsert
\hsize=4.75in
\indent
\boxit{\smallskip
If $\rho = \rho_0$ at $v = v_0$ then $\rho \ge {1\over \rho^{-1}+v_0-v}$.
Thus there is a conjugate point before $v=v_0+\rho^{-1}$.
\smallskip}
\hsize=5.4in
\endinsert

\midinsert
\hskip 0.7in
\epsfbox{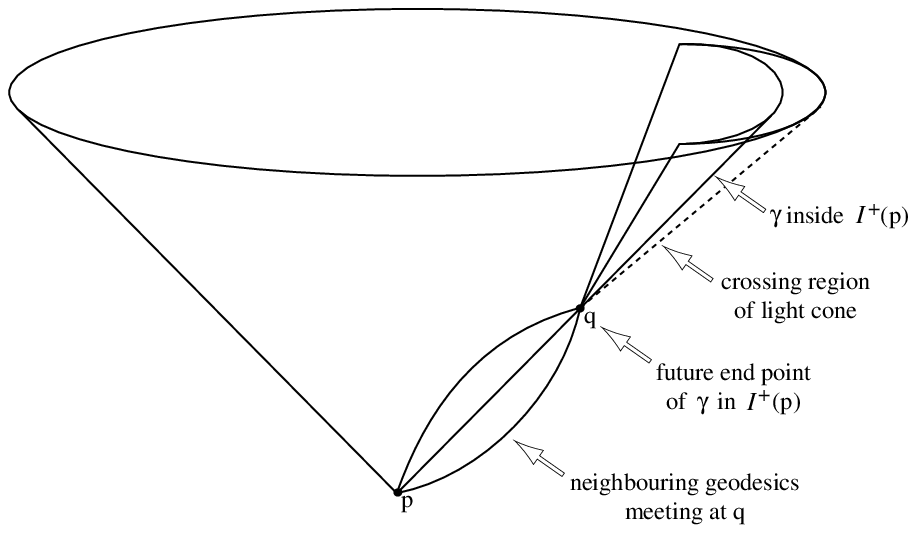}
\medskip
\endinsert

\noindent Infinitesimally neighbouring null
geodesics from $p$ will intersect at $q$. This means the point $q$ will be
conjugate to $p$ along the null geodesic $\gamma$ joining them. For points on
$\gamma$ beyond the conjugate point $q$ there will be a variation of $\gamma$
that
gives a time like curve from $p$. Thus $\gamma$ can not lie in the boundary of
the
future of $p$ beyond the conjugate point $q$. So $\gamma$ will have a future
end
point as a generator of the boundary of the future of $p$.

The situation with time like geodesics is similar, except that the strong
energy
condition that is required to make $R_{ab}l^al^b$ non negative for every time
like vector $l^a$ is, as its name suggests, rather stronger. It is still
however
physically reasonable, at least in an averaged sense, in classical theory. If
the strong energy condition holds, and the time like geodesics from $p$ begin
converging again, then there will be a point $q$ conjugate to $p$.

\midinsert
\hsize=4.75in
\indent
\boxit{\smallskip
{\bf Strong Energy Condition}
$$T_{ab}v^av^b ~~ \ge ~~ {1\over 2}v^av_aT$$
}
\hsize=5.4in
\endinsert

Finally there is the generic energy condition. This says that first the strong
energy condition holds. Second, every time like or null geodesic encounters
some point where there is some curvature that is not specially aligned with the
geodesic. The generic energy condition is not satisfied by a number of known
exact solutions. But these are rather special. One would expect it to be
satisfied by a solution that was "generic" in an appropriate sense. If the
generic energy condition holds, each geodesic will encounter a region of
gravitational focussing. This will imply that there are pairs of conjugate
points if one can extend the geodesic far enough in each direction.

\midinsert
\hsize=4.75in
\indent
\boxit{\smallskip
{\bf The Generic Energy Condition}
\medskip
\item{1.}The strong energy condition holds.
\medskip
\item{2.}Every timelike or null geodesic contains a point
where $l_{[a}R_{b]cd[e}l_{f]}l^cl^d \ne 0$.
\smallskip
}
\hsize=5.4in
\endinsert

One normally thinks of a spacetime singularity as a region in which the
curvature becomes unboundedly large. However, the trouble with that as a
definition is that one could simply leave out the singular points and say that
the remaining manifold was the whole of spacetime. It is therefore better to
define spacetime as the maximal manifold on which the metric is suitably
smooth. One can then recognize the occurrence of singularities by the existence
of incomplete geodesics that can not be extended to infinite values of the
affine parameter.

\midinsert
\hsize=4.75in
\indent
\boxit{\smallskip
{\bf Definition of Singularity}
\medskip
\indent{A spacetime is singular if it is timelike or null geodesically
incomplete, but can not be embedded in a larger spacetime.}\smallskip
}
\hsize=5.4in
\endinsert

\noindent This definition reflects the most objectionable feature of
singularities, that there can be particles whose history has a begining or end
at a finite time. There are examples in which geodesic incompleteness can occur
with the curvature remaining bounded, but it is thought that generically the
curvature will diverge along incomplete geodesics. This is important if one is
to appeal to quantum effects to solve the problems raised by singularities in
classical general relativity.

Between 1965 and 1970 Penrose and I used the techniques I have described to
prove a number of singularity theorems. These theorems had three kinds of
conditions. First there was an energy condition such as the weak, strong or
generic energy conditions. Then there was some global condition on the causal
structure such as that there shouldn't be any closed time like curves. And
finally, there was some condition that gravity was so strong in some region
that nothing could escape.

\midinsert
\hsize=4.75in
\indent
\boxit{\smallskip
{\bf Singularity Theorems}
\medskip
\item{1.}Energy condition.
\smallskip
\item{2.}Condition on global structure.
\smallskip
\item{3.}Gravity strong enough to trap a region.
\smallskip
}
\hsize=5.4in
\endinsert

\noindent This third condition could be expressed in various ways.

\midinsert
\hskip 0.5in
\epsfbox{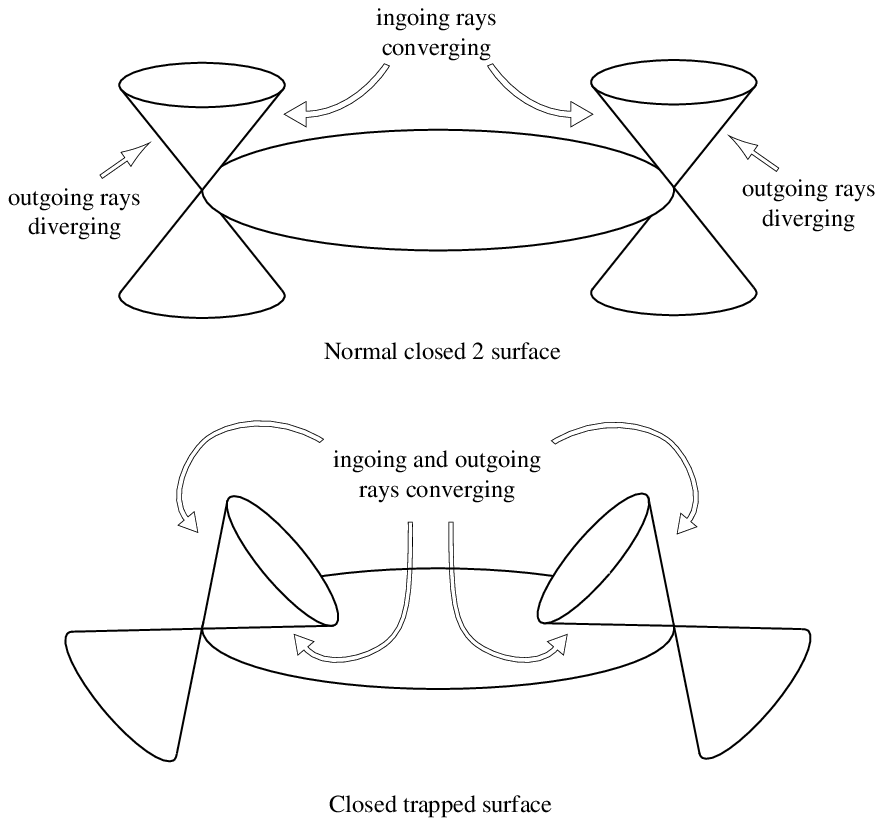}
\endinsert

\noindent One way
would be that the spatial cross section of the universe was closed, for then
there was no outside region to escape to. Another was that there was what was
called a closed trapped surface. This is a closed two surface such that both
the ingoing and out going null geodesics orthogonal to it were converging.
Normally if you have a spherical two surface in Minkowski space the ingoing
null geodesics are converging but the outgoing ones are diverging. But in the
collapse of a star the gravitational field can be so strong that the light
cones are tipped inwards. This means that even the out going null geodesics are
converging.

The various singularity theorems show that spacetime must be time like or null
geodesically incomplete if different combinations of the three kinds of
conditions hold. One can weaken one condition if one assumes stronger versions
of the other two. I shall illustrate this by describing the Hawking-Penrose
theorem. This has the generic energy condition, the strongest of the three
energy conditions. The global condition is fairly weak, that there should be no
closed time like curves. And the no escape condition is the most general, that
there should be either a trapped surface or a closed space like three surface.

\midinsert
\hskip 1.1in
\epsfbox{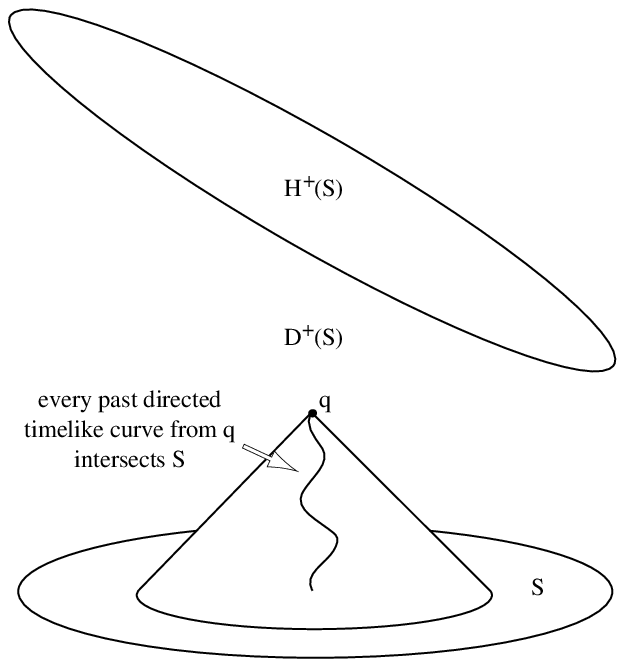}
\endinsert

For simplicity, I shall just sketch the proof  for the case of a closed space
like three surface $S$. One can define the future Cauchy development $D^+(S)$
to
be the region of points $q$ from which every past directed time like curve
intersects $S$. The Cauchy development is the region of spacetime that can be
predicted from data on $S$. Now suppose that the future Cauchy development was
compact. This would imply that the Cauchy development would have a future
boundary called the Cauchy horizon, $H^+(S)$. By an argument similar to that
for
the boundary of the future of a point the Cauchy horizon will be generated by
null geodesic segments without past end points.

\midinsert
\hskip 0.7in
\epsfbox{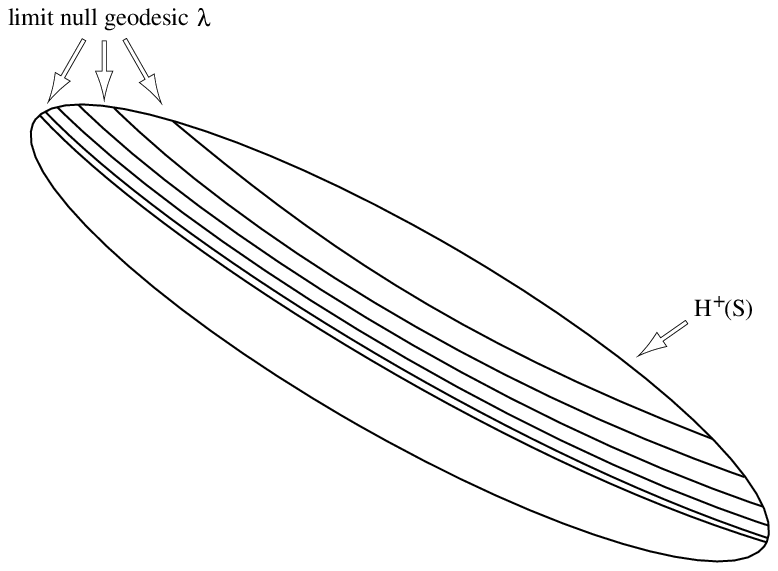}
\endinsert

\noindent However, since the Cauchy
development is assumed to be compact, the Cauchy horizon will also be compact.
This means that the null geodesic generators will wind round and round inside a
compact set. They will approach a limit null geodesic $\lambda$ that will have
no
past or future end points in the Cauchy horizon. But if $\lambda$ were
geodesically complete the generic energy condition would imply that it would
contain conjugate points $p$ and $q$. Points on $\lambda$ beyond $p$ and $q$
could be
joined by a time like curve. But this would be a contradiction because no two
points of the Cauchy horizon can be time like separated. Therefore either
$\lambda$ is not geodesically complete and
the theorem is proved or the future Cauchy development of $S$ is not compact.

\midinsert
\hskip 1.2in
\epsfbox{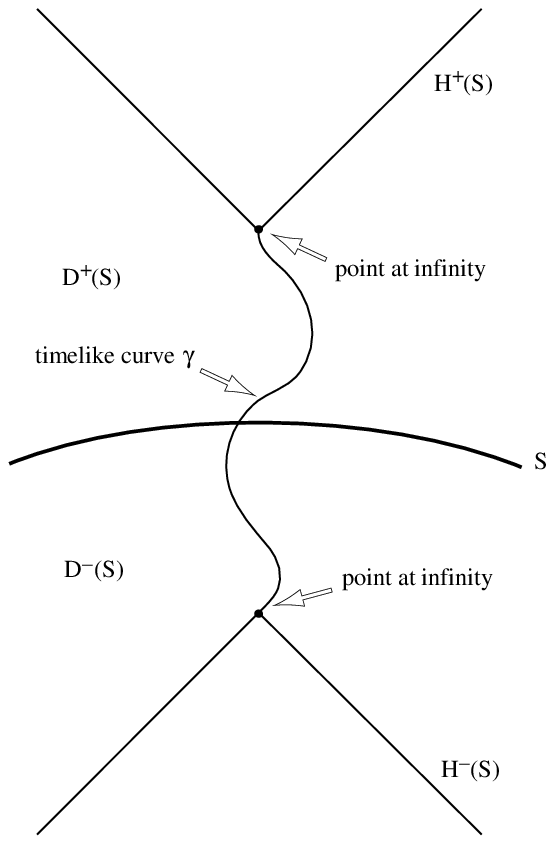}
\endinsert

In the latter case one can show there is a future directed time like curve,
$\gamma$ from $S$ that never leaves the future Cauchy development of $S$. A
rather
similar argument shows that $\gamma$ can be extended to the past to a curve
that
never leaves the past Cauchy development $D^-(S)$.

\midinsert
\hskip 1.2in
\epsfbox{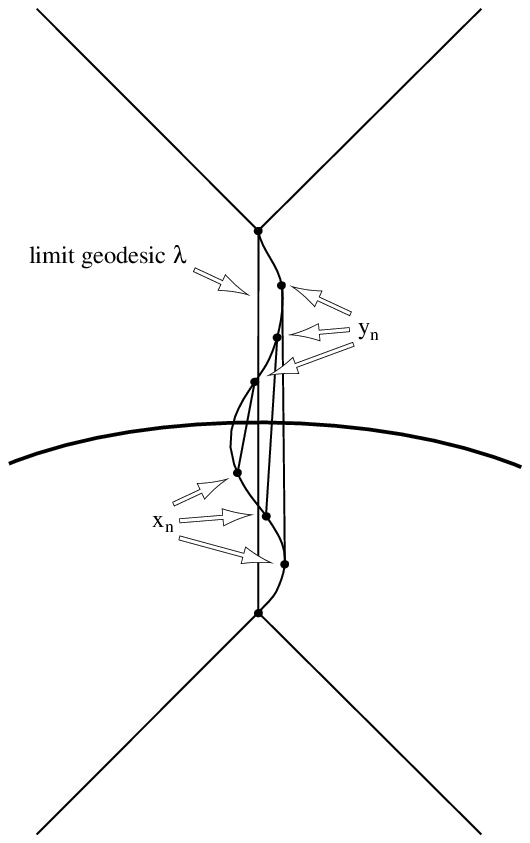}
\endinsert

\noindent Now consider a sequence
of point $x_n$ on $\gamma$ tending to the past and a similar sequence $y_n$
tending to the future. For each  value of $n$ the points $x_n$ and $y_n$ are
time
like separated and are in the globally hyperbolic Cauchy development of $S$.
Thus
there is a time like geodesic of maximum length $\lambda_n$ from $x_n$ to
$y_n$. All
the $\lambda_n$ will cross the compact space like surface $S$. This means that
there
will be a time like geodesic $\lambda$ in the Cauchy development which is a
limit of the time like geodesics $\lambda_n$. Either $\lambda$ will be
incomplete, in
which case the theorem is proved. Or it will contain conjugate poin because
of the generic energy condition. But in that case $\lambda_n$ would contain
conjugate points for $n$ sufficiently large. This would be a contradiction
because the $\lambda_n$ are supposed to be curves of maximum length. One can
therefore conclude that the spacetime is time like or null geodesically
incomplete. In other words there is a singularity.

The theorems predict singularities in two situations. One is in the future in
the gravitational collapse of stars and other massive bodies. Such
singularities would be an end of time, at least for particles moving on the
incomplete geodesics. The other situation in which singularities are predicted
is in the past at the begining of the present expansion of the universe. This
led to the abandonment of attempts (mainly by the Russians) to argue that there
was a previous contracting phase and a non singular bounce into expansion.
Instead almost everyone now believes that the universe, and time itself, had a
begining at the Big Bang. This is a discovery far more important than a few
miscellaneous unstable particles but not one that has been so well recognized
by Nobel prizes.

The prediction of singularities means that classical general relativity is not
a complete theory. Because the singular points have to be cut out of the
spacetime manifold one can not define the field equations there and can not
predict what will come out of a singularity. With the singularity in the past
the only way to deal with this problem seems to be to appeal to quantum
gravity. I shall return to this in my third lecture. But the singularities that
are predicted in the future seem to have a property that Penrose has called,
Cosmic Censorship. That is they conveniently occur in places like black holes
that are hidden from external observers. So any break down of predictability
that may occur at these singularities won't affect what happens in the outside
world, at least not according to classical theory.

\midinsert
\hsize=3.7in
\indent\indent\indent
\boxit{\smallskip
\centerline{\bf Cosmic Censorship}
\medskip
\centerline{Nature abhors a naked singularity}
\smallskip
}
\hsize=5.4in
\endinsert

\noindent However, as I shall show in
the next lecture, there is unpredictability in the quantum theory. This is
related to the fact that gravitational fields can have intrinsic entropy which
is not just the result of coarse graining. Gravitational entropy, and the fact
that time has a begining and may have an end, are the two themes of my
lectures because they are the ways in which gravity is distinctly different
from other physical fields.

The fact that gravity has a quantity that behaves like entropy was first
noticed in the purely classical theory. It depends on Penrose's Cosmic
Censorship Conjecture. This is unproved but is believed to be true for
suitably general initial data and equations of state. I shall use a weak form
of Cosmic Censorship.

\midinsert
\hskip 0.5in
\epsfbox{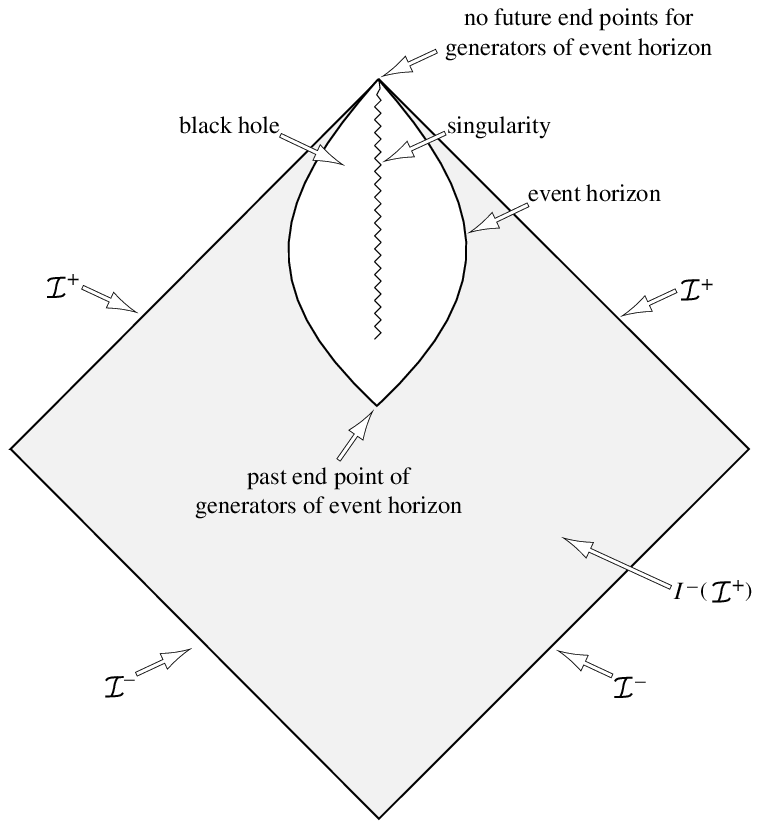}
\endinsert

\noindent One makes the approximation of treating the region around
a collapsing star as asymptotically flat. Then, as Penrose showed, one can
conformally embed the spacetime manifold $M$ in a manifold with boundary ${\bar
M}$.
The boundary $\partial M$ will be a null surface and will consist of two
components, future and past null infinity, called ${\cal I}^+$ and ${\cal
I}^-$. I
shall say that weak Cosmic Censorship holds if two conditions are satisfied.
First, it is assumed that the null geodesic generators of ${\cal I}^+$ are
complete in a certain conformal metric. This implies that observers far from
the
collapse live to an old age and are not wiped out by a thunderbolt
singularity sent out from the collapsing star. Second, it is assumed that the
past of ${\cal I}^+$ is globally hyperbolic. This means there are no naked
singularities that can be seen from large distances. Penrose has a stronger
form of Cosmic Censorship which assumes that the whole spacetime is globally
hyperbolic. But the weak form will suffice for my purposes.

\midinsert
\hsize=4.75in
\indent
\boxit{\smallskip
{\bf Weak Cosmic Censorship}
\medskip
\item{1.}${\cal I}^+$ and ${\cal I}^-$ are complete.
\smallskip
\item{2.}$I^-({\cal I}^+)$ is globally hyperbolic.
\smallskip
}
\hsize=5.4in
\endinsert

If weak Cosmic Censorship holds the singularities that are predicted to occur
in gravitational collapse can't be visible from ${\cal I}^+$. This means that
there must be a region of spacetime that is not in the past of ${\cal I}^+$.
This
region is said to be a black hole because no light or anything else can
escape from it to infinity. The boundary of the black hole region is called the
event horizon. Because it is also the boundary of the past of ${\cal I}^+$ the
event horizon will be generated by null geodesic segments that may have past
end points but don't have any future end points. It then follows that if the
weak energy condition holds the generators of the horizon can't be converging.
For if they were they would intersect each other within a finite distance.

This implies that the area of a cross section of the event horizon can never
decrease with time and in general will increase. Moreover if two black holes
collide and merge together the area of the final black hole will be greater
than the sum of the areas of the original black holes.

\midinsert
\hskip 0.05in
\epsfbox{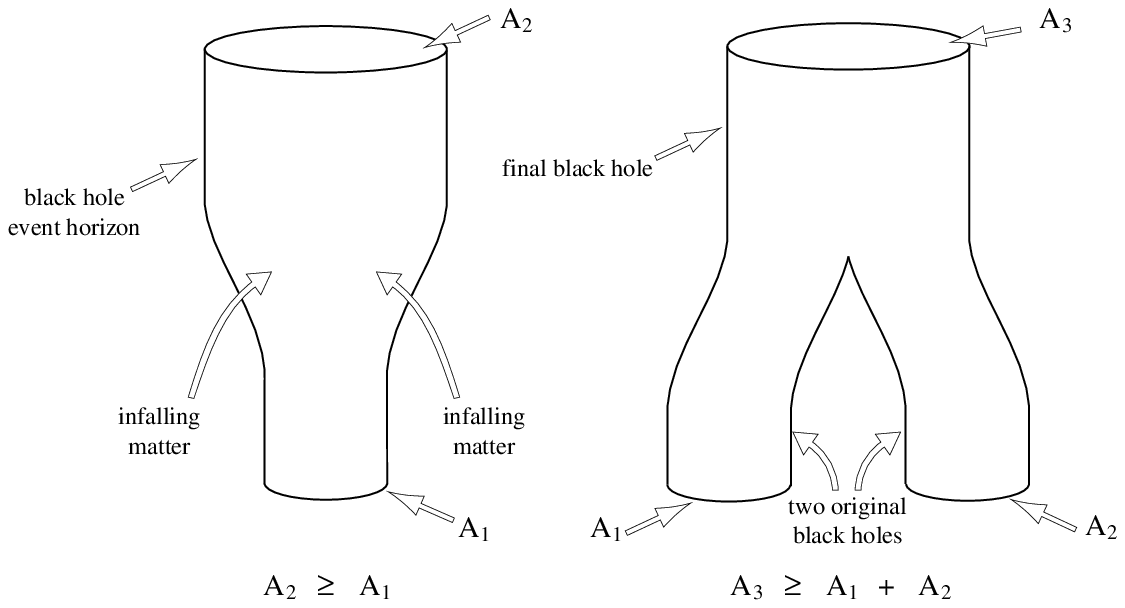}
\endinsert

\noindent This is very similar to
the behavior of entropy according to the Second Law of Thermodynamics. Entropy
can never decrease and the entropy of a total system is greater than the sum
of its constituent parts.

\midinsert
\hsize=4.75in
\indent
\boxit{\smallskip
{\bf Second Law of Black Hole Mechanics}\smallskip
\indent\indent$\delta A ~~ \ge ~~ 0$
\smallskip
{\bf Second Law of Thermodynamics}\smallskip
\indent\indent$\delta S ~~ \ge ~~ 0$
\smallskip
}
\hsize=5.4in
\endinsert

The similarity with thermodynamics is increased by what is called the First
Law of Black Hole Mechanics. This relates the change in mass of a black hole to
the change in the area of the event horizon and the change in its angular
momentum and electric charge. One can compare this to the First Law of
Thermodynamics which gives the change in internal energy in terms of the
change in entropy and the external work done on the system.

\midinsert
\hsize=4.75in
\indent
\boxit{\smallskip
{\bf First Law of Black Hole Mechanics}
$$\delta E ~~ = ~~ {\kappa\over 8\pi}\delta A ~~ + ~~
\Omega\delta J ~~ + ~~ \Phi\delta Q$$
\medskip
{\bf First Law of Thermodynamics}
$$\delta E ~~ = ~~ T\delta S ~~ + ~~ P\delta V$$
\smallskip
}
\hsize=5.4in
\endinsert

\noindent One sees that if
the area of the event horizon is analogous to entropy then the quantity
analogous to temperature is what is called the surface gravity of the black
hole $\kappa$. This is a measure of the strength of the gravitational field on
the
event horizon. The similarity with thermodynamics is further increased by the
so called Zeroth Law of Black Hole Mechanics: the surface gravity is the same
everywhere on the event horizon of a time independent black hole.

\midinsert
\hsize=4.75in
\indent
\boxit{\smallskip
{\bf Zeroth Law of Black Hole Mechanics}
\smallskip
\indent\indent $\kappa$ is the same everywhere on the horizon of a time
independent
\smallskip\indent black hole.
\medskip
{\bf Zeroth Law of Thermodynamics}
\smallskip
\indent\indent $T$ is the same everywhere for a system in thermal equilibrium.
\medskip
}
\hsize=5.4in
\endinsert

Encouraged by these similarities Bekenstein proposed that some multiple of the
area of the event horizon actually was the entropy of a black hole. He
suggested a generalized Second Law: the sum of this black hole entropy and the
entropy of matter outside black holes would never decrease.

\midinsert
\hsize=4.75in
\indent
\boxit{\smallskip
{\bf Generalised Second Law}
$$\delta (S + c A) ~~ \ge ~~ 0$$
\smallskip
}
\hsize=5.4in
\endinsert

\noindent However this
proposal was not consistent. If black holes have an entropy proportional to
horizon area they should also have a non zero temperature proportional to
surface gravity.

\midinsert
\hskip 1.0in
\epsfbox{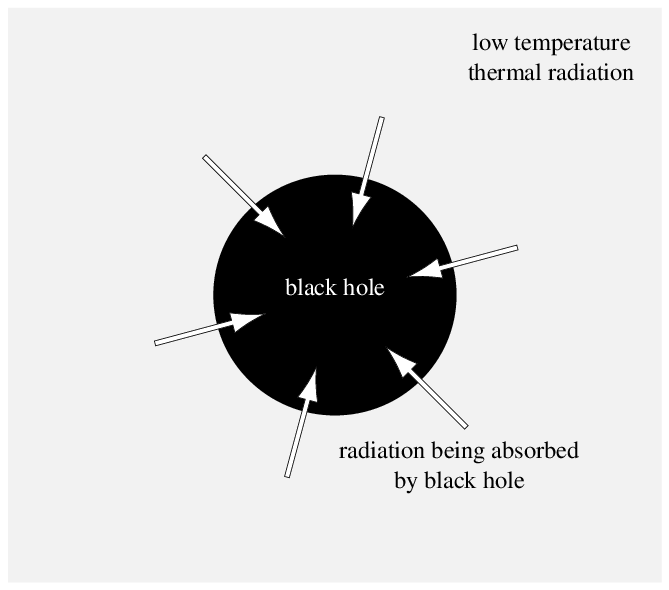}
\endinsert

\noindent Consider a black hole that is in contact with thermal
radiation at a temperature lower than the black hole temperature. The black
hole will absorb some of the radiation but won't be able to send anything out,
because according to classical theory nothing can get out of a black hole. One
thus has heat flow from the low temperature thermal radiation to the higher
temperature black hole. This would violate the generalized Second Law because
the loss of entropy from the thermal radiation would be greater than the
increase in black hole entropy. However, as we shall see in my next lecture,
consistency was restored when it was discovered that black holes are sending
out radiation that was exactly thermal. This is too beautiful a result to be a
coincidence or just an approximation. So it seems that black holes really do
have intrinsic gravitational entropy. As I shall show, this is related to the
non trivial topology of a black hole. The intrinsic entropy means that gravity
introduces an extra level of unpredictability over and above the uncertainty
usually associated with quantum theory. So Einstein was wrong when he said
``God
does not play dice''. Consideration of black holes suggests, not only that God
does play dice, but that He sometimes confuses us by throwing them where they
can't be seen.
\vfill\eject

\midinsert
\epsfbox{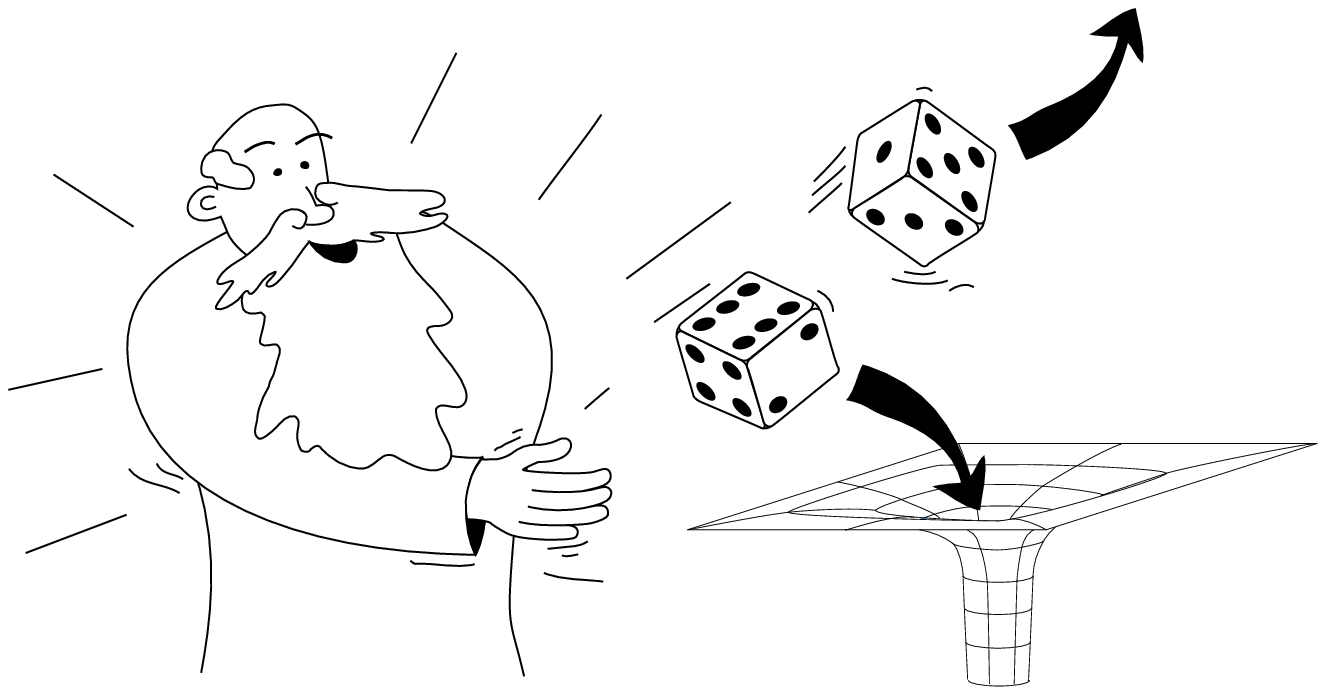}
\endinsert

\vfill\eject


\centerline{\bf 2. Quantum Black Holes}
\smallskip
\centerline{S. W. Hawking}
\bigskip

In my second lecture I'm going to talk about the quantum theory of black holes.
It seems to lead to a new level of unpredictability in physics over and above
the usual uncertainty associated with quantum mechanics. This is because black
holes appear to have intrinsic entropy and to lose information from our region
of the universe. I should say that these claims are controversial: many people
working on quantum gravity, including almost all those that entered it from
particle physics, would instinctively reject the idea that information about
the
quantum state of a system could be lost. However they have had very little
success in showing how information can get out of a black hole. Eventually I
believe they will be forced to accept my suggestion that it is lost, just as
they were forced to agree that black holes radiate, which was against all their
preconceptions.

I should start by reminding you about the classical theory of black holes. We
saw in the last lecture that gravity is always attractive, at least in normal
situations. If gravity had been sometimes attractive and sometimes repulsive,
like electro-dynamics, we would never notice it at all because it is about $10
^{40}$ times weaker. It is only because gravity always has the same sign
that the gravitational force between the particles of two macroscopic bodies
like ourselves and the Earth add up to give a force we can feel.

The fact that gravity is attractive means that it will tend to draw the matter
in the universe together to form objects like stars and galaxies. These can
support themselves for a time against further contraction by thermal pressure,
in the case of stars, or by rotation and internal motions, in the case of
galaxies. However, eventually the heat or the angular momentum will be
carried away and the object will begin to shrink. If the mass is less than
about one and a half times that of the Sun the contraction can be stopped by
the degeneracy pressure of electrons or neutrons. The object will settle down
to
be a white dwarf or a neutron star respectively. However, if the mass is
greater than this limit there is nothing that can hold it up and stop it
continuing to contract. Once it has shrunk to a certain critical size the
gravitational field at its surface will be so strong that the light cones will
be bent inward as in the diagram on the following page. I would have liked to
draw you a four
dimensional picture. However, government cuts have meant that Cambridge
university can afford only two dimensional screens. I have therefore shown
time in the vertical direction and used perspective to show two of the three
space directions. You can see that even the outgoing light rays are bent
towards
each other and so are converging rather than diverging. This means that there
is a closed trapped surface which is one of the alternative third conditions of
the Hawking-Penrose theorem.

\midinsert
\hskip 0.6in
\epsfbox{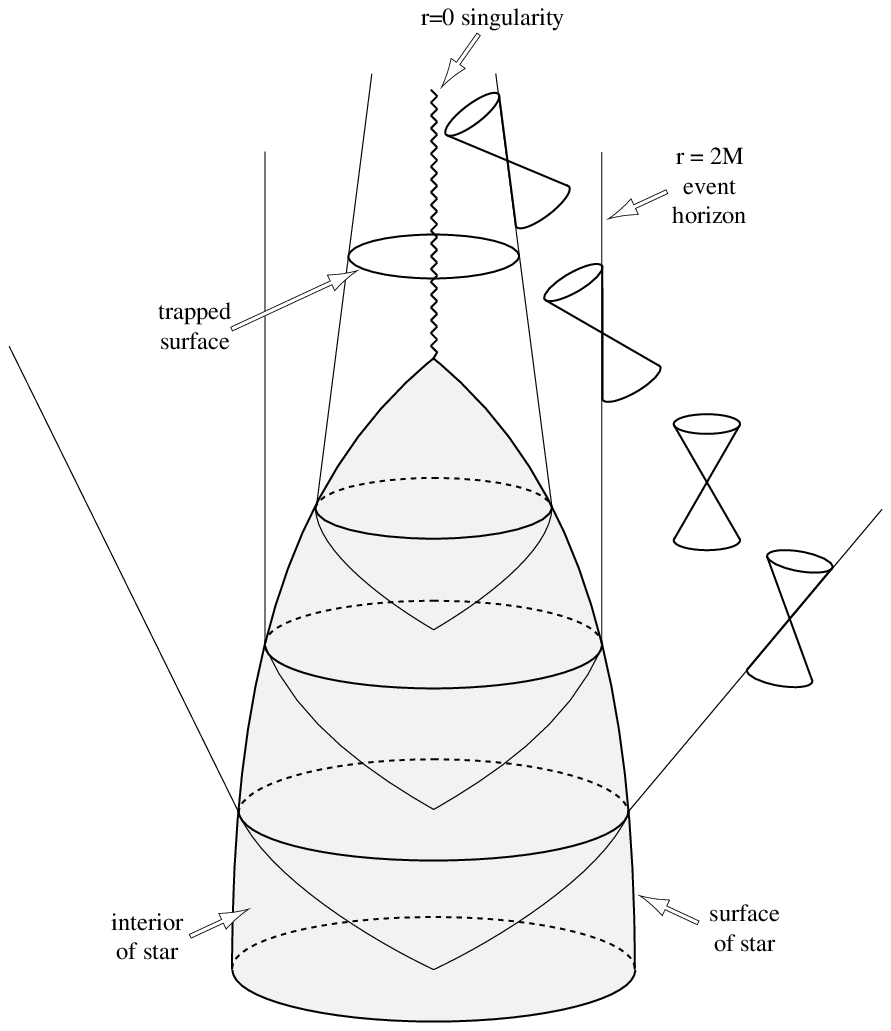}
\endinsert

If the Cosmic Censorship Conjecture is correct the trapped surface and the
singularity it predicts can not be visible from far away. Thus there must be a
region of spacetime from which it is not possible to escape to infinity. This
region is said to be a black hole. Its boundary is called the event horizon and
it is a null surface formed by the light rays that just fail to get away to
infinity. As we saw in the last lecture, the area of a cross section of the
event horizon can never decrease, at least in the classical theory.  This, and
perturbation calculations of spherical collapse, suggest that black holes will
settle down to a stationary state. The no hair theorem, proved by the combined
work of Israel, Carter, Robinson and myself, shows that the only stationary
black holes in the absence of matter fields are the Kerr solutions. These are
characterized by two parameters, the mass $M$ and the angular momentum $J$. The
no hair theorem was extended by Robinson to the case where there was an
electromagnetic field. This added a third parameter $Q$, the electric charge.
The
no hair theorem has not been proved for the Yang-Mills field, but the only
difference seems to be the addition of one or more integers that label a
discrete family of unstable solutions. It can be shown that there are no more
continuous degrees of freedom of time independent Einstein-Yang-Mills black
holes.

\midinsert
\hsize=4.0in
\indent\indent\boxit{
\hskip 0.5in
\epsfbox{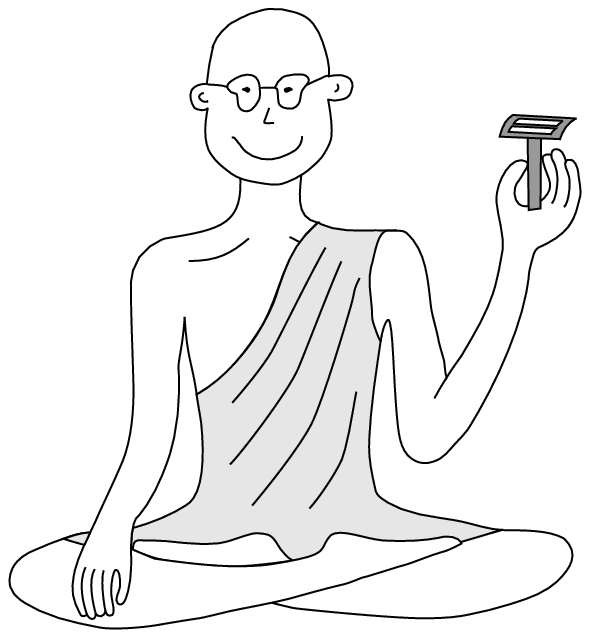}
\smallskip
{\bf No Hair Theorem}
\smallskip
\indent Stationary black holes are characterised by mass $M$,
angular momentum $J$ and electric charge $Q$.\smallskip}
\hsize=5.4in
\endinsert

What the no hair theorems show is that a large amount of information is lost
when a body collapses to form a black hole. The collapsing body is described by
a very large number of parameters. There are the types of matter and the
multipole
moments of the mass distribution. Yet the black hole that forms is
completely independent of the type of matter and rapidly loses all the
multipole
moments except the first two: the monopole moment, which is the mass, and
the dipole moment, which is the angular momentum.

This loss of information didn't really matter in the classical theory. One
could say that all the information about the collapsing body was still inside
the black hole. It would be very difficult for an observer outside the black
hole to determine what the collapsing body was like. However, in the classical
theory it was still possible in principle. The observer would never actually
lose sight of the collapsing body. Instead it would appear to slow down and
get very dim as it approached the event horizon. But the observer could still
see what it was made of and how the mass was distributed. However, quantum
theory changed all this. First, the collapsing body would send out only a
limited number of photons before it crossed the event horizon. They would be
quite insufficient to carry all the information about the collapsing body. This
means that in quantum theory there's no way an outside observer can measure
the state of the collapsed body. One might not think this mattered too much
because the information would still be inside the black hole even if one
couldn't measure it from the outside. But this is where the second effect of
quantum theory on black holes comes in. As I will show, quantum theory will
cause black holes to radiate and lose mass. Eventually it seems that they will
disappear completely, taking with them the information inside them. I will give
arguments that this information really is lost and doesn't come back in some
form. As I will show, this loss of information would introduce a new level of
uncertainty into physics over and above the usual uncertainty associated with
quantum theory. Unfortunately, unlike Heisenberg's Uncertainty Principle, this
extra level will be rather difficult to confirm experimentally in the case of
black holes. But as I will argue in my third lecture, there's a sense in which
we may have already observed it in the measurements of fluctuations in the
microwave background.

The fact that quantum theory causes black holes to radiate was first discovered
by doing quantum field theory on the background of a black hole formed by
collapse. To see how this comes about it is helpful to use what are normally
called Penrose diagrams. However, I think Penrose himself would agree they
really should be called Carter diagrams because Carter was the first to use
them systematically. In a spherical collapse the spacetime won't depend on the
angles $\theta$ and $\phi$. All the geometry will take place in the $r$-$t$
plane.
Because any two dimensional plane is conformal to flat space one can represent
the causal structure by a diagram in which null lines in the $r$-$t$ plane are
at
$\pm 45$ degrees to the vertical.

\midinsert
\hskip 1.2in
\epsfbox{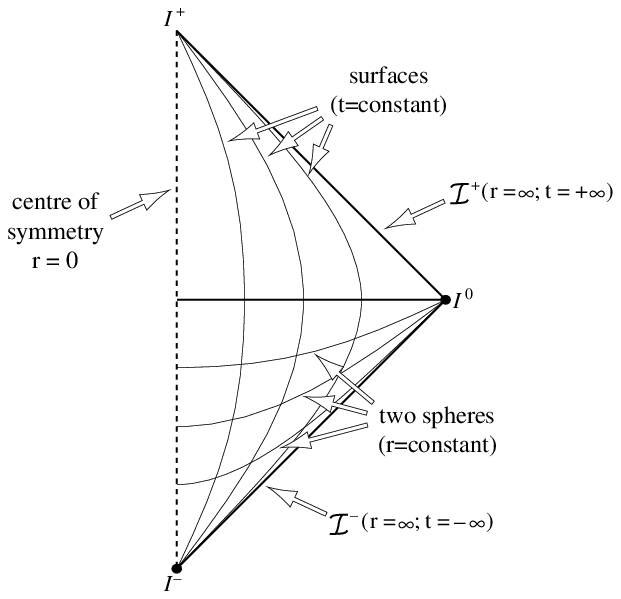}
\endinsert

\noindent Let's start with flat Minkowski
space. That has a Carter-Penrose diagram which is a triangle standing on one
corner. The two diagonal sides on the right correspond to the past and future
null infinities I referred to in my first lecture. These are really at
infinity but all distances are shrunk by a conformal factor as one approaches
past or future null infinity. Each point of this triangle corresponds to a two
sphere of radius $r$. $r=0$ on the vertical line on the left, which represents
the
center of symmetry, and $r\rightarrow\infty$ on the right of the diagram.

One can easily see from the diagram that every point in Minkowski space is in
the past of future null infinity ${\cal I}^+$. This means there is no black
hole
and no event horizon. However, if one has a spherical body collapsing the
diagram is rather different.

\midinsert
\hskip 1.2in
\epsfbox{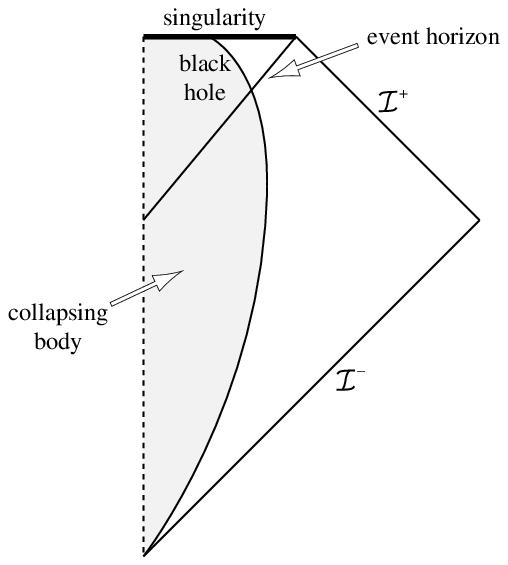}
\endinsert

\noindent It looks the same in the past but now the top of
the triangle has been cut off and replaced by a horizontal boundary. This is
the singularity that the Hawking-Penrose theorem predicts. One can now see that
there are points under this horizontal line that are not in the past of future
null infinity ${\cal I}^+$. In other words there is a black hole. The event
horizon, the boundary of the black hole, is a diagonal line that comes down
from
the top right corner and meets the vertical line corresponding to the center of
symmetry.

One can consider a scalar field $\phi$ on this background. If the spacetime
were
time independent, a solution of the wave equation, that contained only positive
frequencies on scri minus, would also be positive frequency on scri plus. This
would mean that there would be no particle creation, and there would be no out
going particles on scri plus, if there were no scalar particles initially.

However, the metric is time dependent during the collapse. This will cause a
solution that is positive frequency on ${\cal I}^-$ to be partly negative
frequency when it gets to ${\cal I}^+$. One can calculate this mixing by taking
a
wave with time dependence $e^{-i \omega u}$ on ${\cal I}^+$ and
propagating it back to ${\cal I}^-$. When one does that one finds that the part
of the wave that passes near the horizon is very blue shifted. Remarkably it
turns out that the mixing is independent of the details of the collapse in the
limit of late times. It depends only on the surface gravity $\kappa$ that
measures the strength of the gravitational field on the horizon of the black
hole. The mixing of positive and negative frequencies leads to particle
creation.

When I first studied this effect in 1973 I expected I would find a burst of
emission during the collapse but that then the particle creation would die out
and one would be left with a black hole that was truely black. To my great
surprise I found that after a burst during the collapse there remained a
steady rate of particle creation and emission. Moreover, the emission was
exactly thermal with a temperature of ${\kappa \over 2\pi}$. This was just what
was
required to make consistent the idea that a black hole had an entropy
proportional to the area of its event horizon. Moreover, it fixed the constant
of proportionality to be a quarter in Planck units, in which $G=c ={\bar h}
=1$.
This makes the unit of area $10^{-66}~{\rm cm}^2$ so a black
hole of the mass of the Sun would have an entropy of the order of $10^{78}$.
This would reflect the enormous number of different ways in which it could be
made.

\midinsert
\hsize=4.0in
\indent\indent
\boxit{\smallskip
\centerline{\bf Black Hole Thermal Radiation}
$${\rm Temperature} ~ T ~~ = ~~ {\kappa\over 2\pi}$$
$${\rm Entropy} ~ S ~~ = ~~ {1\over 4} A$$}
\hsize=5.4in
\endinsert

When I made my original discovery of radiation from black holes it seemed a
miracle that a rather messy calculation should lead to emission that was
exactly thermal. However, joint work with Jim Hartle and Gary Gibbons uncovered
the deep reason. To explain it I shall start with the example of the
Schwarzschild metric.

\midinsert
\hsize=4.75in
\indent
\boxit{\smallskip
\centerline{\bf Schwarzschild Metric}
$$ds^2 ~ = - \left(1-{2M\over r}\right) dt^2 ~+~
\left(1-{2M\over r}\right)^{\hskip -3pt -1}
\hskip-7pt dr^2
{}~+~r^2(d\theta^2+\sin^2\theta d\phi^2)$$}
\hsize=5.4in
\endinsert

\noindent This represents the gravitational field that a black hole
would settle down to if it were non rotating. In the usual $r$ and $t$
coordinates
there is an apparent singularity at the Schwarzschild radius $r=2M$. However,
this is just caused by a bad choice of coordinates. One can choose other
coordinates in which the metric is regular there.

\midinsert
\hskip 0.8in
\epsfbox{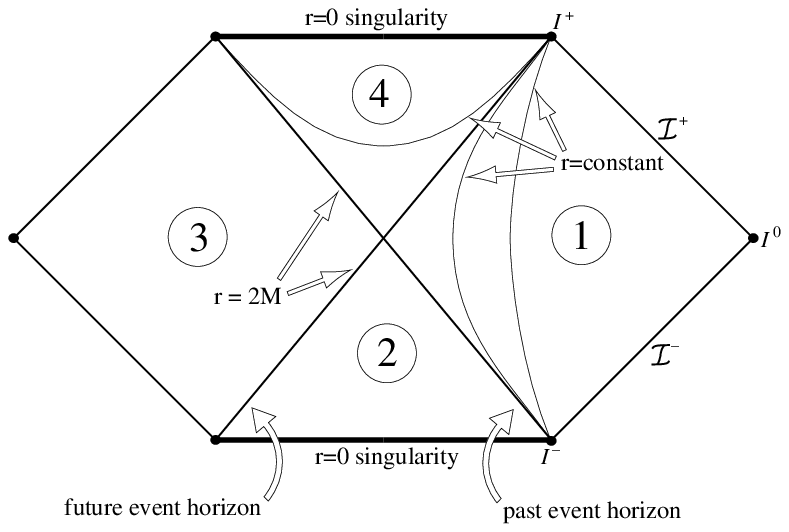}
\endinsert

\noindent The Carter-Penrose diagram has
the form of a diamond with flattened top and bottom. It is divided into four
regions by the two null surfaces on which $r=2M$. The region on the right,
marked
$\bigcirc$\hskip -7pt {\eightrm 1}\hskip 7pt on
the diagram is the asymptotically flat space in which we are supposed to
live. It has past and future null infinities ${\cal I}^-$ and ${\cal I}^+$ like
flat spacetime.  There is another asymptotically flat region
$\bigcirc$\hskip -7pt {\eightrm 3}\hskip 7pt on the left that
seems to correspond to another universe that is connected to ours only through
a wormhole. However, as we shall see, it is connected to our region through
imaginary time. The null surface from bottom left to top right is the boundary
of the region from which one can escape to the infinity on the right. Thus it
is
the future event horizon. The epithet future being added to distinguish it
from the past event horizon which goes from bottom right to top left.

Let us now return to the Schwarzschild metric in the original $r$ and $t$
coordinates. If one puts $t=i \tau$ one gets a positive definite metric. I
shall
refer to such positive definite metrics as Euclidean even though they may be
curved. In the Euclidean-Schwarzschild metric there is again an apparent
singularity at $r=2M$. However, one can define a new radial coordinate $x$ to
be
$4M(1 -2Mr^{-1})^{1\over 2}$.

\midinsert
\hsize=4.75in
\indent
\boxit{\smallskip
\centerline{\bf Euclidean-Schwarzschild Metric}
$$ds^2 ~ = ~ x^2\left({d\tau\over 4M}\right)^{\hskip-3pt 2} + \hskip 2pt
\left({r^2\over 4M^2}\right)^{\hskip-3pt 2}dx^2 ~+~
r^2(d\theta^2+\sin^2\theta d\phi^2)$$}
\hsize=5.4in
\endinsert

\noindent The metric in the $x-\tau$ plane
then becomes like the origin of polar coordinates if one identifies the
coordinate $\tau$ with period $8 \pi M$. Similarly other Euclidean black hole
metrics will have apparent singularities on their horizons which can be
removed by identifying the imaginary time coordinate with period ${2\pi\over
\kappa}$.

\midinsert
\hskip 1.4in
\epsfbox{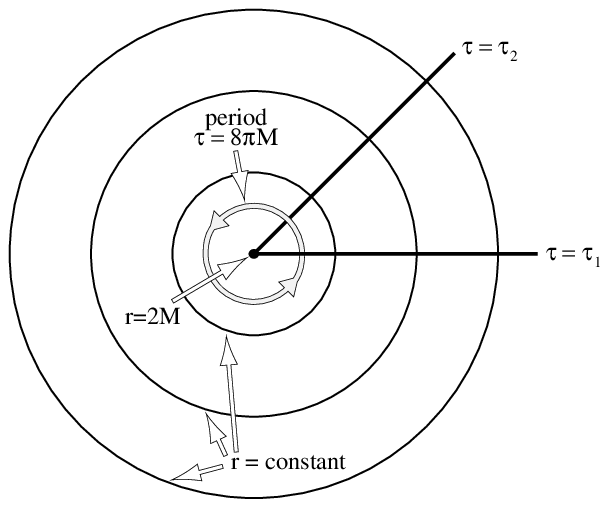}
\endinsert

So what is the significance of having imaginary time identified with some
period $\beta$. To see this consider the amplitude to go from some field
configuration $\phi_1$ on the surface $t_1$ to a configuration $\phi_2$ on the
surface $t_2$. This will be given by the matrix element of $e^{iH(t_2-t_1)}$.
 However, one can also represent this amplitude
as a path integral over all fields $\phi$ between $t_1$ and $t_2$ which agree
with the
given fields $\phi_1$ and $\phi_2$ on the two surfaces.

\midinsert
\hskip 1.3in
\epsfbox{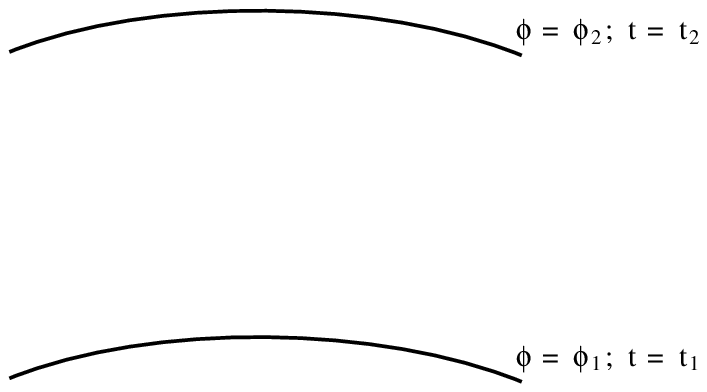}
\smallskip
$$\eqalign{<\phi_2,t_2~|~\phi_1,t_1> ~& = ~
<\phi_2~|~\exp (-iH(t_2-t_1))~|~\phi_1>\cr
{}~& = ~ \int D[\phi ] \exp (iI[\phi])\cr}$$
\endinsert

One now chooses the time separation $(t_2-t_1)$ to be pure imaginary and
equal to $\beta$. One also puts the initial field $\phi_1$ equal to the final
field $\phi_2$ and sums over a complete basis of states $\phi_n$.  On the left
one
has the expectation value of $e^{-\beta H}$ summed
over all states. This is just the thermodynamic partition function $Z$
at the temperature $T = \beta^{-1}$.

\midinsert
\hskip 1.1in
\epsfbox{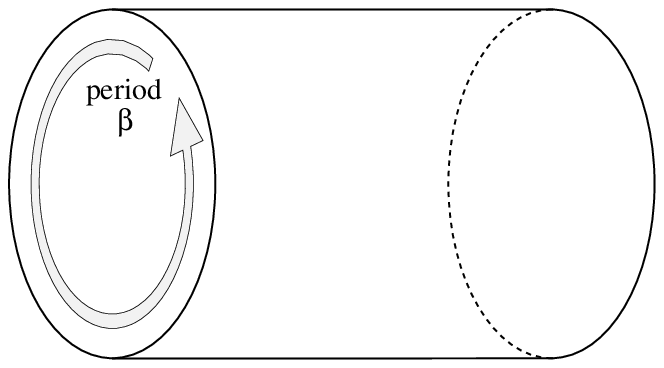}
\smallskip
$$t_2-t_1  =  -i\beta , ~~~ \phi_2 ~ = ~ \phi_1$$
$$\eqalign{Z ~& = ~ \sum <\phi_n ~ | ~ \exp (-\beta H ) ~|~ \phi_n>\cr
&=~\int D[\phi]\exp (-i{\hat I}[\phi ])\cr}$$
\endinsert

On the right hand of the equation one has a path integral. One puts
$\phi_1=\phi
_2$ and sums over all field configurations $\phi_n$. This means that
effectively
one is doing the path integral over all fields $\phi$ on a spacetime that is
identified periodically in the imaginary time direction with period $\beta$.
Thus
the partition function for the field $\phi$ at temperature $T$ is given
by a path integral over all fields on a Euclidean spacetime. This spacetime is
periodic in the imaginary time direction with period $\beta = T^{-1}$.

If one does the path integral in flat spacetime identified with period $\beta$
in
the imaginary time direction one gets the usual result for the partition
function of black body radiation. However, as we have just seen, the Euclidean-
Schwarzschild solution is also periodic in imaginary time with period ${2\pi
\over \kappa}$. This means that fields on the Schwarzschild background will
behave
as if they were in a thermal state with temperature ${\kappa \over 2 \pi}$.

The periodicity in imaginary time explained why the messy calculation of
frequency mixing led to radiation that was exactly thermal. However, this
derivation avoided the problem of the very high frequencies that take part in
the frequency mixing approach. It can also be applied when there are
interactions between the quantum fields on the background. The fact that the
path integral is on a periodic background implies that all physical quantities
like expectation values will be thermal. This would have been very difficult to
establish in the frequency mixing approach.

One can extend these interactions to include interactions with the
gravitational field itself. One starts with a background metric $g_0$ such as
the
Euclidean-Schwarzschild metric that is a solution of the classical field
equations. One can then expand the action $I$ in a power series in the
perturbations $\delta g$ about $g_0$.

$$I[g] ~ = ~ I[g_0] ~+~ I_2(\delta g)^2 ~+~ I_3 (\delta g)^3 ~+~ ...$$

\noindent The linear term vanishes because the
background is a solution of the field equations. The quadratic term can be
regarded as describing gravitons on the background while the cubic and higher
terms describe interactions between the gravitons. The path integral over the
quadratic terms are finite. There are non renormalizable divergences at two
loops in pure gravity but these cancel with the fermions in supergravity
theories. It is not known whether supergravity theories have divergences at
three loops or higher because no one has been brave or foolhardy enough to
try the calculation. Some recent work indicates that they may be finite to all
orders. But even if there are higher loop divergences they will make very
little difference except when the background is curved on the scale of the
Planck length, $10^{-33}$ cm.

More interesting than the higher order terms is the zeroth order term, the
action of the background metric $g_0$.

$$I=-{1\over 16\pi}\int R(-g)^{1\over 2}~ d^4x
{}~+~ {1\over 8\pi}\int K(\pm h)^{1\over 2}~d^3x$$

\noindent The usual Einstein-Hilbert action for
general relativity is the volume integral of the scalar curvature $R$.
This is zero for vacuum solutions so one might think that the action of the
Euclidean-Schwarzschild solution was zero. However, there is also a surface
term in the action proportional to the integral of $K$, the trace of the
second fundemental form of the boundary surface. When one includes this and
subtracts off the surface term for flat space one finds the action of the
Euclidean-Schwarzschild metric is ${\beta^2\over 16\pi}$ where $\beta$ is the
period in imaginary time at infinity. Thus the dominant contribution to the
path
integral for the partition function $Z$ is $e^{-\beta^2\over 16\pi}$.

$$Z~=~\sum\exp (-\beta E_n)~=~\exp \left(-{\beta^2\over 16\pi}\right)$$

If one differentiates $\log Z$ with respect to the period $\beta$ one gets the
expectation value of the energy, or in other words, the mass.

$$<E> = - {d\over d\beta}(\log Z) ~ = ~ {\beta\over 8\pi}$$

\noindent So this gives the
mass $M ={\beta \over 8 \pi}$. This confirms the relation between the mass and
the
period, or inverse temperature, that we already knew. However, one can go
further. By standard thermodynamic arguments, the log of the partition function
is equal to minus the free energy $F$ divided by the temperature
$T$.

$$\log Z = -{F\over T}$$

\noindent And the free energy is the mass or energy plus the temperature
times the entropy $S$.

$$F ~=~ <E> ~+~ TS$$

\noindent Putting all this together one sees that the
action of the black hole gives an entropy of $4 \pi M^2$.

$$S~=~{\beta^2\over 16\pi} ~=~ 4\pi M^2 ~=~ {1\over 4} A$$

\noindent This is exactly what is required to make the laws
of black holes the same as the laws of thermodynamics.

Why does one get this intrinsic gravitational entropy which has no parallel in
other quantum field theories. The reason is  gravity allows different
topologies for the spacetime manifold.

\midinsert
\hskip 1.4in
\epsfbox{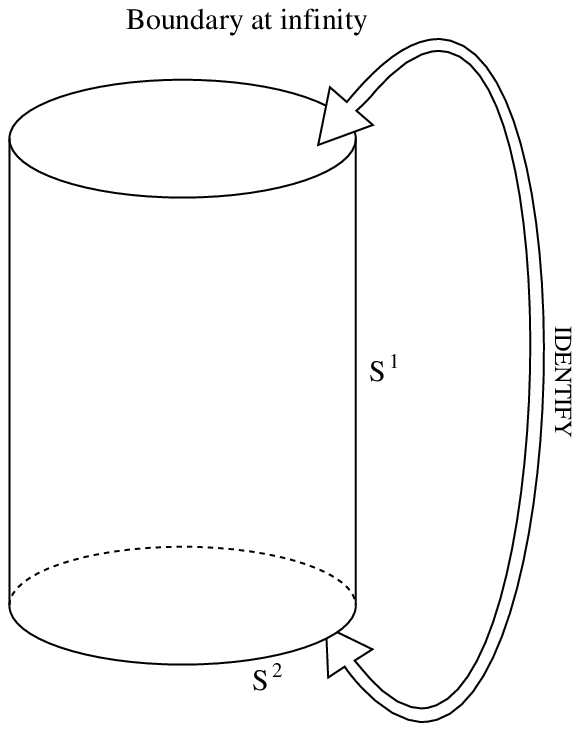}
\endinsert

\noindent In the case we are considering the
Euclidean-Schwarzschild solution has a boundary at infinity that has topology
$S^2 \times S^1$. The $S^2$ is a large space like two sphere at infinity and
the $S^1$
corresponds to the imaginary time direction which is identified periodically.
One can fill in this boundary with metrics of at least two different
topologies. One of course is the Euclidean-Schwarzschild metric. This has
topology $R^2 \times S^2$, that is the Euclidean two plane times a two sphere.
The
other is $R^3 \times S^1$, the topology of Euclidean flat space periodically
identified in the imaginary time direction. These two topologies have different
Euler numbers. The Euler number of periodically identified flat space is zero,
while that of the Euclidean-Schwarzschild solution is two.

\midinsert
\hskip 0.6in
\epsfbox{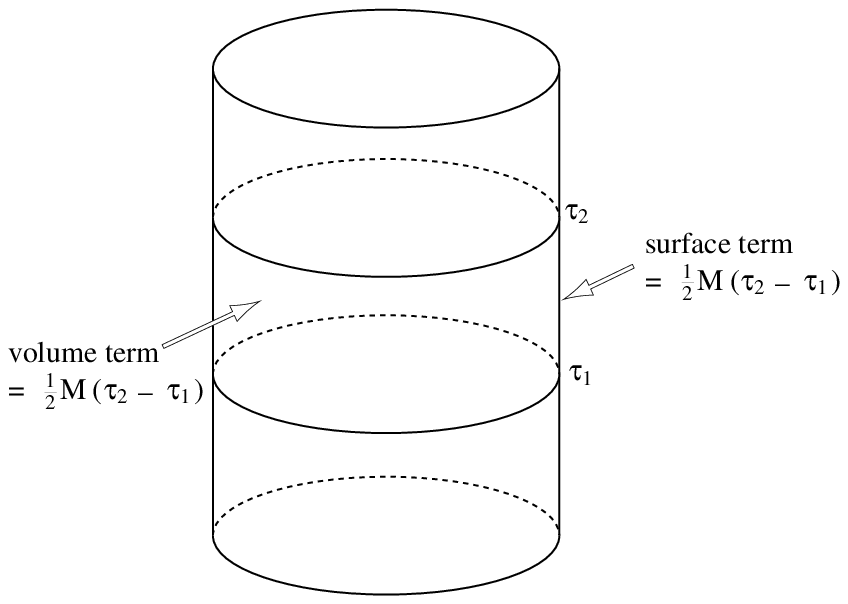}
$${\rm Total~action} = M(\tau_2-\tau_1)$$
\endinsert

\noindent The significance of
this is as follows: on the topology of periodically identified flat space one
can find a periodic time function $\tau$ whose gradient is no where zero and
which agrees with the imaginary time coordinate on the boundary at infinity.
One can then work out the action of the region between two surfaces $\tau_1$
and
$\tau_2$. There will be two contributions to the action, a volume integral over
the
matter Lagrangian, plus the Einstein-Hilbert Lagrangian and a surface term. If
the solution is time independent the surface term over $\tau = \tau_1$ will
cancel
with the surface term over $\tau = \tau_2$. Thus the only net contribution to
the
surface term comes from the boundary at infinity. This gives half the mass
times the imaginary time interval $(\tau_2 - \tau_1)$. If the mass is non-zero
there must be non-zero matter fields to create the mass. One can show that the
volume integral over the matter Lagrangian plus the Einstein-Hilbert
Lagrangian also gives ${1\over 2}M (\tau_2 -\tau_1)$. Thus the total
action is $M (\tau_2 -\tau_1)$. If one puts this contribution to
the log of the partition function into the thermodynamic formulae one finds
the expectation value of the energy to be the mass, as one would expect.
However, the entropy contributed by the background field will be zero.

The situation is different however with the Euclidean-Schwarzschild solution.

\midinsert
\hskip 1.3in
\epsfbox{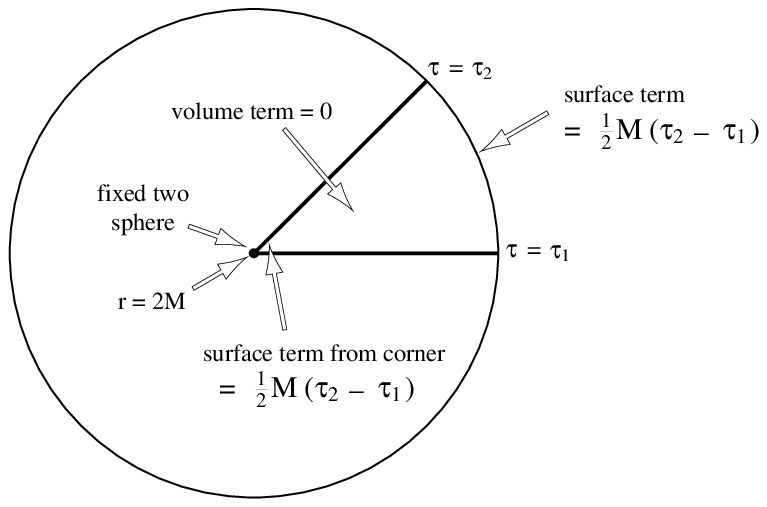}
\vskip 0.01in
$${\rm Total~action~including~corner~contribution} = M(\tau_2-\tau_1)$$
$${\rm Total~action~without~corner~contribution} = {1\over 2}M(\tau_2-\tau_1)$$
\endinsert

\noindent Because the Euler number is two rather than zero one can't find a
time
function $\tau$ whose gradient is everywhere non-zero. The best one can do is
choose the imaginary time coordinate of the Schwarzschild solution. This has a
fixed two sphere at the horizon where $\tau$ behaves like an angular
coordinate.
If one now works out the action between two surfaces of constant $\tau$ the
volume
integral vanishes because there are no matter fields and the scalar curvature
is zero. The trace $K$ surface term at infinity again gives ${1\over 2}M
(\tau_2 -\tau_1)$.
However there is now another surface term at the
horizon where the $\tau_1$ and $\tau_2$ surfaces meet in a corner. One can
evaluate
this surface term and find that it also is equal to ${1\over 2}M (\tau_2
-\tau_1)$.
Thus the total action for the region between $\tau_1$ and $\tau_2$ is
$M (\tau_2 -\tau_1)$. If one used this action with $\tau_2 -\tau_1= \beta$
one would find that the entropy was zero. However, when one looks at
the action of the Euclidean Schwarzschild solution from a four dimensional
point of view rather than a 3+1, there is no reason to include a surface term
on the horizon because the metric is regular there. Leaving out the surface
term on the horizon reduces the action by one quarter the area of the horizon,
which is just the intrinsic gravitational entropy of the black hole.

The fact that the entropy of black holes is connected with a topological
invariant, the Euler number, is a strong argument that it will remain even if
we have to go to a more fundemental theory. This idea is anathema to most
particle physicists who are a very conservative lot and want to make
everything like Yang-Mills theory. They agree that the radiation from black
holes seems to be thermal and independent of how the hole was formed if the
hole is large compared to the Planck length. But they would claim that when the
black hole loses mass and gets down to the Planck size, quantum general
relativity will break down and all bets will be off. However, I shall describe
a thought experiment with black holes in which information seems to be lost
yet the curvature outside the horizons always remains small.

It has been known for some time that one can create pairs of positively and
negatively charged particles in a strong electric field. One way of looking at
this is to note that in flat Euclidean space a particle of charge $q$ such as
an electron would move in a circle in a uniform electric field $E$.
One can analytically continue this motion from the imaginary time $\tau$ to
real
time $t$. One gets a pair of positively and negatively charged particles
accelerating away from each other pulled apart by the electric field.

\midinsert
\hskip 0.8in
\epsfbox{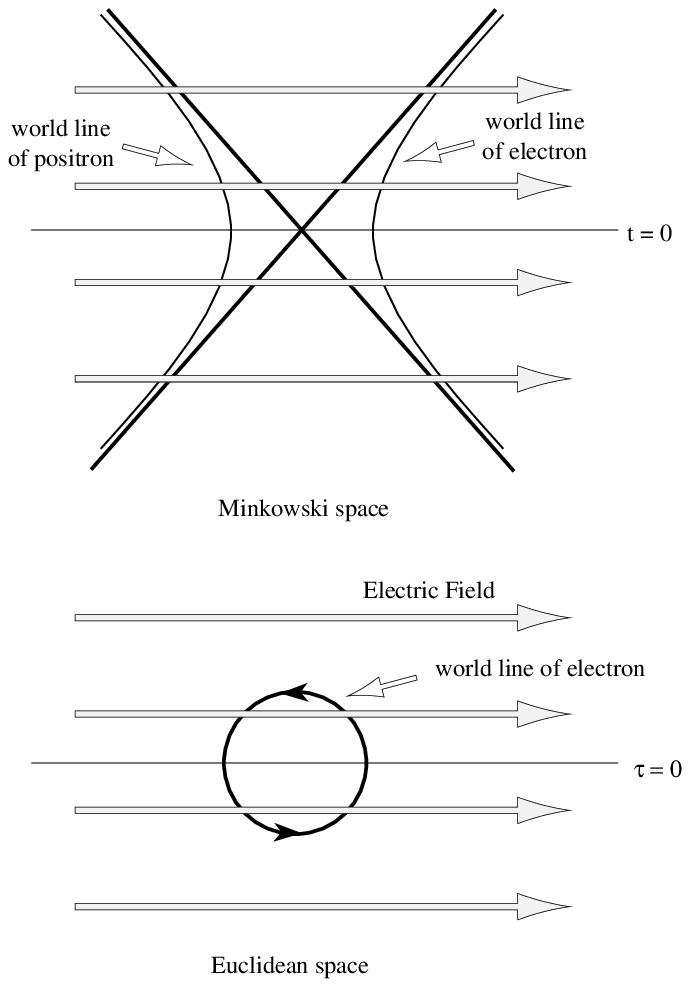}
\endinsert

The process of pair creation is described by chopping the two diagrams in half
along the $t=0$ or $\tau =0$ lines. One then joins the upper half of the
Minkowski
space diagram to the lower half of the Euclidean space diagram.

\midinsert
\hskip 0.8in
\epsfbox{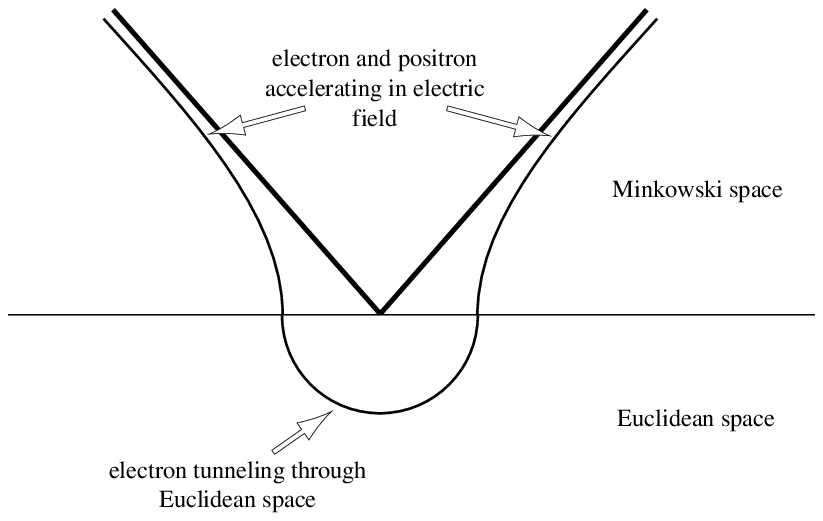}
\endinsert

\noindent This gives a
picture in which the positively and negatively charged particles are really the
same particle. It tunnels through Euclidean space to get from one Minkowski
space world line to the other. To a first approximation the probability for
pair creation is $e^{-I}$ where
$$
{\rm Euclidean~ action~} I ~ = ~{2 \pi m^2\over q E}.
$$
Pair creation by strong electric fields has been observed
experimentally and the rate agrees with these estimates.

Black holes can also carry electric charges so one might expect that they could
also be pair created. However the rate would be tiny compared to that for
electron positron pairs because the mass to charge ratio is $10^{20}$ times
bigger. This means that any electric field would be neutralized by electron
positron pair creation long before there was a significant probability of pair
creating black holes. However there are also black hole solutions with
magnetic charges. Such black holes couldn't be produced by gravitational
collapse because there are no magnetically charged elementary particles. But
one might expect that they could be pair created in a strong magnetic field. In
this case there would be no competition from ordinary particle creation
because ordinary particles do not carry magnetic charges. So the magnetic field
could become strong enough that there was a significant chance of creating a
pair of magnetically charged black holes.

In 1976 Ernst found a solution that represented two magnetically charged black
holes accelerating away from each other in a magnetic field.

\midinsert
\hskip 1.2in
\epsfbox{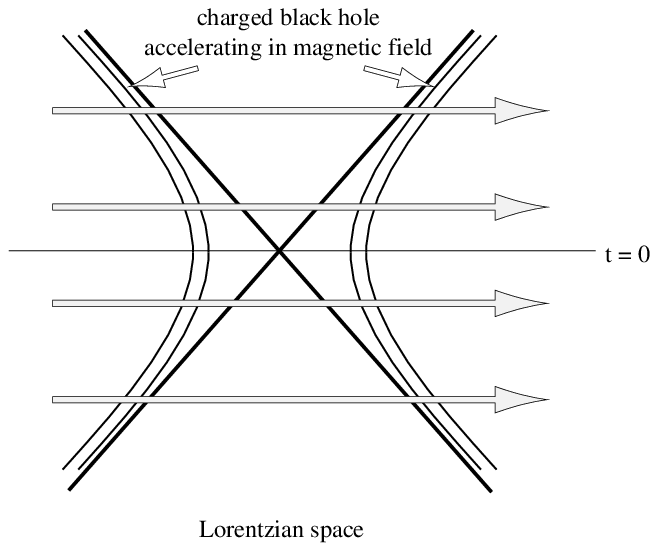}
\endinsert

\midinsert
\hskip 1.2in
\epsfbox{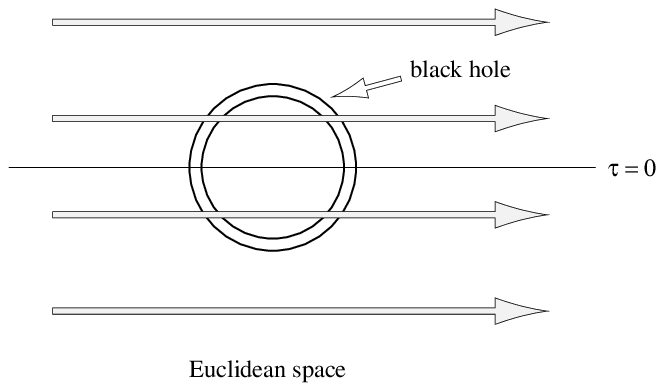}
\endinsert

\noindent If one
analytically continues it to imaginary time one has a picture very like that of
the electron pair creation. The black hole moves on a circle in a curved
Euclidean space just like the electron moves in a circle in flat Euclidean
space. There is a complication in the black hole case because the imaginary
time coordinate is periodic about the horizon of the black hole as well as
about the center of the circle on which the black hole moves. One has to adjust
the mass to charge ratio of the black hole to make these periods equal.
Physically this means that one chooses the parameters of the black hole so
that the temperature of the black hole is equal to the temperature it sees
because it is accelerating.. The temperature of a magnetically charged black
hole tends to zero as the charge tends to the mass in Planck units. Thus for
weak magnetic fields, and hence low acceleration, one can always match the
periods.

Like in the case of pair creation of electrons one can describe pair creation
of black holes by joining the lower half of the imaginary time Euclidean
solution to the upper half of the real time Lorentzian solution.

\midinsert
\hskip 0.8in
\epsfbox{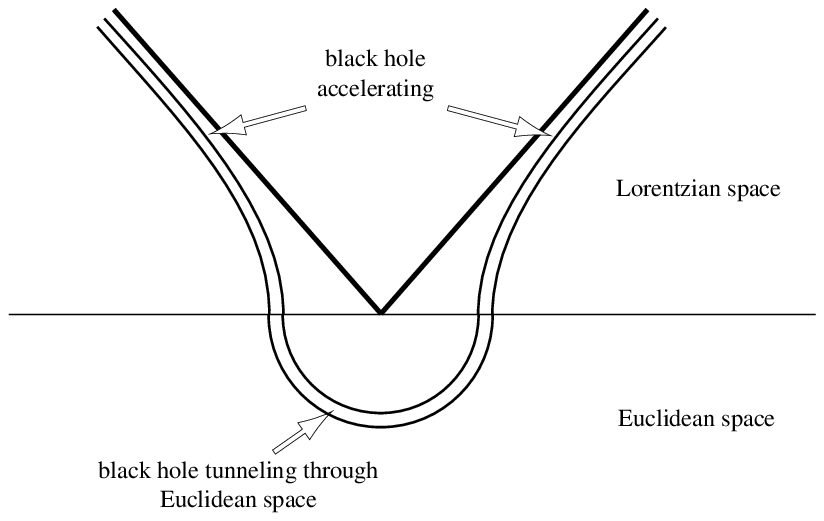}
\endinsert

\noindent One can think of the black hole as tunneling through the Euclidean
region and emerging as a
pair of oppositely charged black holes that accelerate away from each other
pulled apart by the magnetic field. The accelerating black hole solution is not
asymptotically flat because it tends to a uniform magnetic field at infinity.
But one can nevertheless use it to estimate the rate of pair creation of black
holes in a local region of magnetic field.

\midinsert
\hskip 0.9in
\epsfbox{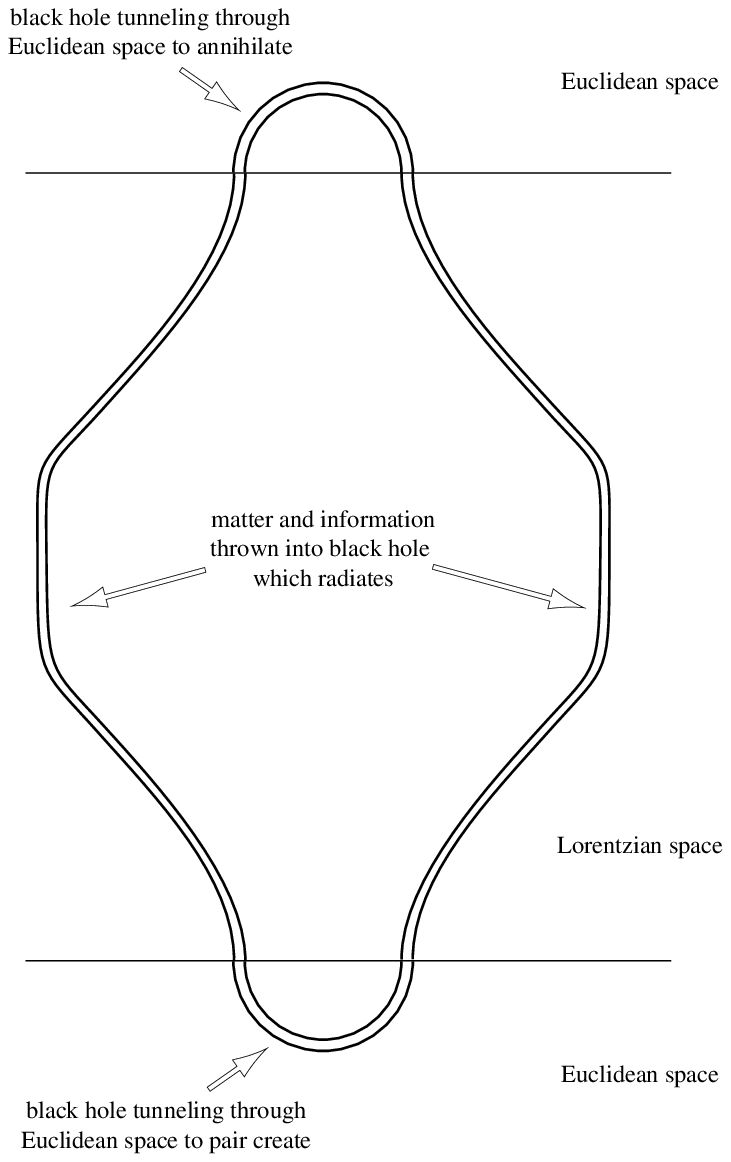}
\endinsert

\noindent One could imagine that after being
created the black holes move far apart into regions without magnetic field.
One could then treat each black hole separately as a black hole in
asymptotically flat space. One could throw an arbitrarily large amount of
matter
and information into each hole. The holes would then radiate and lose mass.
However, they couldn't lose magnetic charge because there are no magnetically
charged particles. Thus they would eventually get back to their original state
with the mass slightly bigger than the charge. One could then bring the two
holes back together again and let them annihilate each other. The annihilation
process can be regarded as the time reverse of the pair creation. Thus it is
represented by the top half of the Euclidean solution joined to the bottom half
of the Lorentzian solution. In between the pair creation and the annihilation
one can have a long Lorentzian period in which the black holes move far apart,
accrete matter, radiate and then come back together
again. But the topology of
the gravitational field will be the topology of the Euclidean Ernst solution.
This is $S^2 \times S^2$ minus a point.

One might worry that the Generalized Second Law of Thermodynamics would be
violated when the black holes annihilated because the black hole horizon area
would have disappeared. However it turns out that the area of the acceleration
horizon in the Ernst solution is reduced from the area it would have if there
were no pair creation. This is a rather delicate calculation because the area
of the acceleration horizon is infinite in both cases. Nevertheless there is a
well defined sense in which their difference is finite and equal to the black
hole horizon area plus the difference in the action of the solutions with and
without pair creation. This can be understood as saying that pair creation is a
zero energy process; the Hamiltonian {\it with} pair creation is the same as
the Hamiltonian {\it without}. I'm very grateful to Simon Ross and Gary
Horovitz
for calculating
this reduction just in time for this lecture. It is miracles like this, and I
mean the result not that they got it, that convince me that black hole
thermodynamics can't just be a low energy approximation. I believe that
gravitational entropy won't disappear even if we have to go to a more
fundemental theory of quantum gravity.

One can see from this thought experiment that one gets intrinsic gravitational
entropy and loss of information when the topology of spacetime is different
from that of flat Minkowski space. If the black holes that pair create are
large compared to the Planck size the curvature outside the horizons will be
everywhere small compared to the Planck scale. This means the approximation I
have made of ignoring cubic and higher terms in the perturbations should be
good. Thus the conclusion that information can be lost in black holes should be
reliable.

If information is lost in macroscopic black holes it should also be lost in
processes in which microscopic, virtual black holes appear because of quantum
fluctuations of the metric. One could imagine that particles and information
could fall into these holes and get lost. Maybe that is where all those odd
socks went. Quantities like energy and electric charge, that are coupled to
gauge fields, would be conserved but other information and global charge
would be lost. This would have far reaching implications for quantum theory.

It is normally assumed that a system in a pure quantum state evolves in a
unitary way through a succession of pure quantum states. But if there is loss
of information through the appearance and disappearance of black holes there
can't be a unitary evolution. Instead the loss of information will mean that
the final state after the black holes have disappeared will be what is called
a mixed quantum state. This can be regarded as an ensemble of different pure
quantum states each with its own probability. But because it is not with
certainty in any one state one can not reduce the probability of the final
state to zero by interfering with any quantum state. This means that gravity
introduces a new level of unpredictability into physics over and above the
uncertainty usually associated with quantum theory. I shall show in the next
lecture we may have already observed this extra uncertainty. It means an end to
the hope of scientific determinism that we could predict the future with
certainty. It seems God still has a few tricks up his sleeve.

\midinsert
\hskip 0.5in
\epsfbox{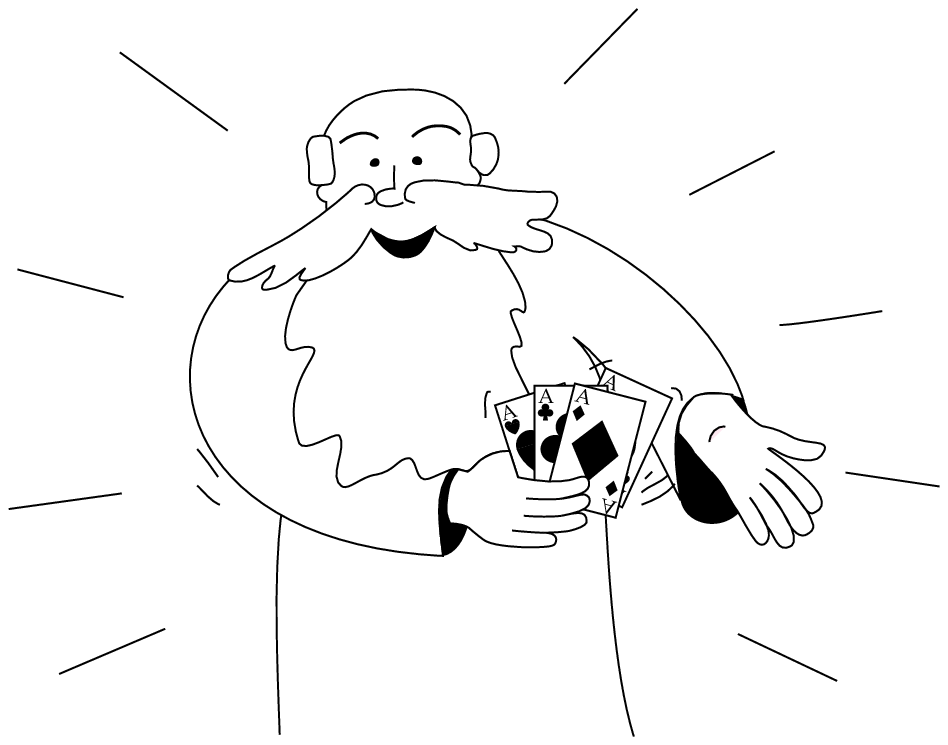}
\endinsert
\vfill\eject


\centerline{\bf 3. Quantum Cosmology}
\smallskip
\centerline{S. W. Hawking}
\bigskip

In my third lecture I shall turn to cosmology. Cosmology used to be considered
a pseudo-science and the preserve of physicists who may have done useful work
in their earlier years but who had gone mystic in their dotage. There were two
reasons for this. The first was that there was an almost total absence of
reliable observations. Indeed, until the 1920s about the only important
cosmological observation was that the sky at night is dark. But people didn't
appreciate the significance of this. However, in recent years the range and
quality of cosmological observations has improved enormously with developments
in technology. So this objection against regarding cosmology as a science, that
it doesn't have an observational basis is no longer valid.

There is, however, a second and more serious objection. Cosmology can not
predict anything about the universe unless it makes some assumption about the
initial conditions. Without such an assumption, all one can say is that things
are as they are now because they were as they were at an earlier stage. Yet
many people believe that science should be concerned only with the local laws
which govern how the universe evolves in time. They would feel that the
boundary conditions for the universe that determine how the universe began
were a question for metaphysics or religion rather than science.

The situation was made worse by the theorems that Roger and I proved. These
showed that according to general relativity there should be a singularity in
our past. At this singularity the field equations could not be defined. Thus
classical general relativity brings about its own downfall: it predicts that it
can't predict the universe.

Although many people welcomed this conclusion, it has always profoundly
disturbed me. If the laws of physics could break down at the begining of the
universe, why couldn't they break down any where. In quantum theory it is a
principle that anything can happen if it is not absolutely forbidden. Once one
allows that singular histories could take part in the path integral they could
occur any where and predictability would disappear completely. If the laws of
physics break down at singularities, they could break down any where.

The only way to have a scientific theory is if the laws of physics hold
everywhere including at the begining of the universe. One can regard this as a
triumph for the principles of democracy: Why should the begining of the
universe be exempt from the laws that apply to other points. If all points are
equal one can't allow some to be more equal than others.

To implement the idea that the laws of physics hold everywhere, one should take
the path integral only over non-singular metrics. One knows in the ordinary
path
integral case that the measure is concentrated on non-differentiable paths. But
these are the completion in some suitable topology of the set of smooth paths
with well defined action. Similarly, one would expect that the path integral
for
quantum gravity should be taken over the completion of the space of smooth
metrics. What the path integral can't include is metrics with singularities
whose action is not defined.

In the case of black holes we saw that the path integral should be taken over
Euclidean, that is, positive definite metrics. This meant that the
singularities
of black holes, like the Schwarzschild solution, did not appear on the
Euclidean
metrics which did not go inside the horizon. Instead the horizon was like the
origin of polar coordinates. The action of the Euclidean metric was therefore
well defined. One could regard this as a quantum version of Cosmic Censorship:
the break down of the structure at a singularity should not affect any physical
measurement.

It seems, therefore, that the path integral for quantum gravity should be taken
over non-singular Euclidean metrics. But what should the boundary conditions be
on these metrics. There are two, and only two, natural choices. The first is
metrics that approach the flat Euclidean metric outside a compact set. The
second possibility is metrics on manifolds that are compact and without
boundary.

\midinsert
\hsize=4.75in
\indent
\boxit{\smallskip
\indent{\bf Natural choices for path integral for quantum gravity}
\smallskip
\item{1.}Asymptotically Euclidean metrics.
\smallskip
\item{2.}Compact metrics without boundary.
\smallskip}
\hsize=5.4in
\endinsert

The first class of asymptotically Euclidean metrics is obviously appropriate
for scattering calculations.

\midinsert
\hskip 1.4in
\epsfbox{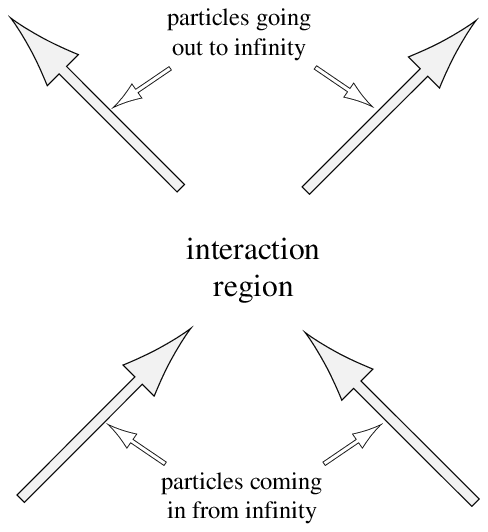}
\endinsert

\noindent In these one sends particles in from infinity and
observes what comes out again to infinity. All measurements are made at
infinity where one has a flat background metric and one can interpret small
fluctuations in the fields as particles in the usual way. One doesn't ask what
happens in the interaction region in the middle. That is why one does a path
integral over all possible histories for the interaction region, that is, over
all asymptotically Euclidean metrics.

However, in cosmology one is interested in measurements that are made in a
finite region rather than at infinity. We are on the inside of the universe
not looking in from the outside. To see what difference this makes let us first
suppose that the path integral for cosmology is to be taken over all
asymptotically Euclidean metrics.

\midinsert
\hskip 0.8in
\epsfbox{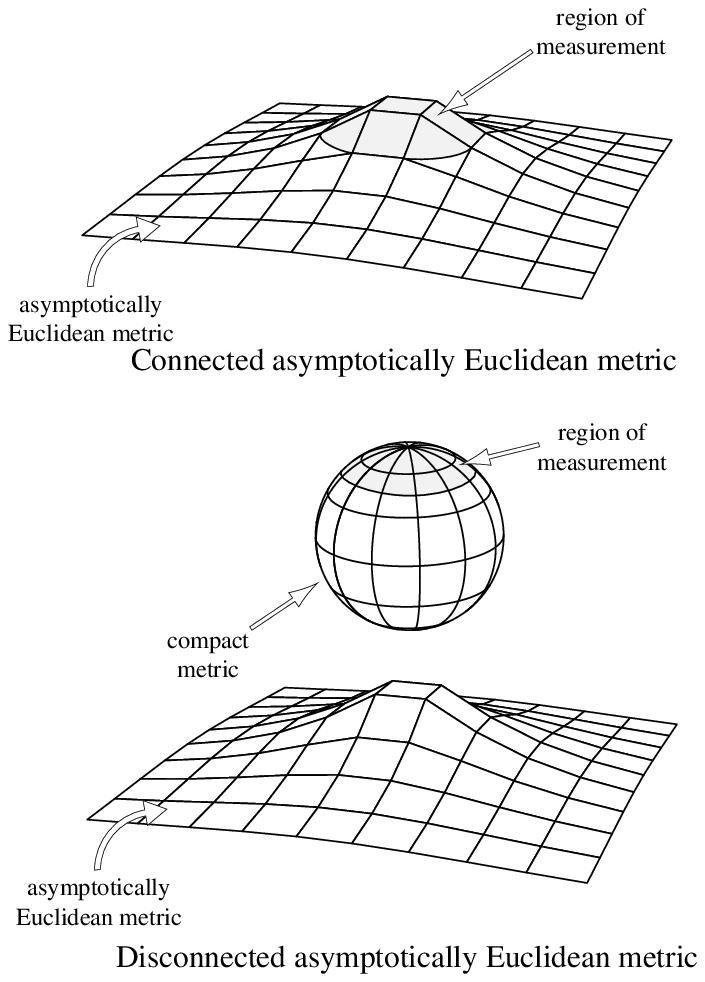}
\endinsert

\noindent Then there would be two contributions to
probabilities for measurements in a finite region. The first would be from
connected asymptotically Euclidean metrics. The second would be from
disconnected metrics that consisted of a compact spacetime containing the
region of measurements and a separate asymptotically Euclidean metric. One can
not exclude disconnected metrics from the path integral because they can be
approximated by connected metrics in which the different components are joined
by thin tubes or wormholes of neglible action.

Disconnected compact regions of spacetime won't affect scattering calculations
because they aren't connected to infinity, where all measurements are made. But
they will affect measurements in cosmology that are made in a finite region.
Indeed, the contributions from such disconnected metrics will dominate over the
contributions from connected asymptotically Euclidean metrics. Thus, even if
one
took the path integral for cosmology to be over all asymptotically Euclidean
metrics, the effect would be almost the same as if the path integral had been
over all compact metrics. It therefore seems more natural to take the path
integral for cosmology to be over all compact metrics without boundary, as Jim
Hartle and I proposed in 1983.

\midinsert
\hsize=4.75in
\indent
\boxit{\smallskip
{\bf The No Boundary Proposal (Hartle and Hawking)}
\medskip
The path integral for quantum gravity should be taken
 over all compact Euclidean metrics.
\smallskip}
\hsize=5.4in
\endinsert

\noindent One can paraphrase this as The Boundary
Condition Of The Universe Is That It Has No Boundary.

In the rest of this lecture I shall show that this no boundary proposal seems
to account for the universe we live in. That is an isotropic and homogeneous
expanding universe with small perturbations. We can observe the spectrum and
statistics of these perturbations in the fluctuations in the microwave
background. The results so far agree with the predictions of the no boundary
proposal. It will be a real test of the proposal and the whole Euclidean
quantum gravity program when the observations of the microwave background are
extended to smaller angular scales.

In order to use the no boundary proposal to make predictions, it is useful to
introduce a concept that can describe the state of the universe at one time.

\midinsert
\hskip 1.6in
\epsfbox{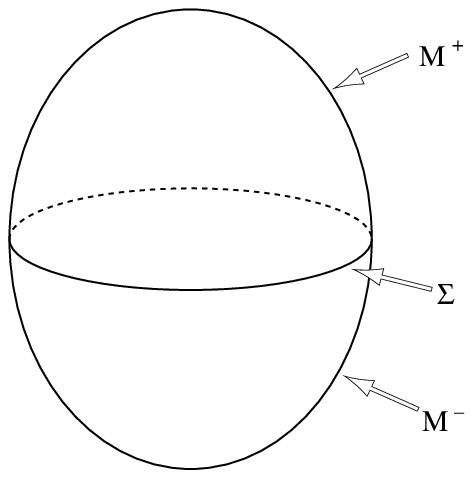}
$${\rm Probability~ of~ induced~ metric~} h_{ij} {\rm ~on~} \Sigma~ =~
\int_{{\rm metrics ~on}~M~{\rm that}\atop{\rm induce}~h_{ij}
{}~{\rm on}~\Sigma}
d[g]~e^{-I}$$
\endinsert

\noindent Consider the probability that the spacetime manifold $M$ contains an
embedded
three dimensional manifold $\Sigma$ with induced metric $h_{ij}$. This is given
by
a path integral over all metrics $g_{ab}$ on $M$ that induce $h_{ij}$ on
$\Sigma$. If $M$
is simply connected, which I will assume, the surface $\Sigma$ will divide $M$
into
two parts $M^+$ and $M^-$.

\midinsert
\hskip 1.6in
\epsfbox{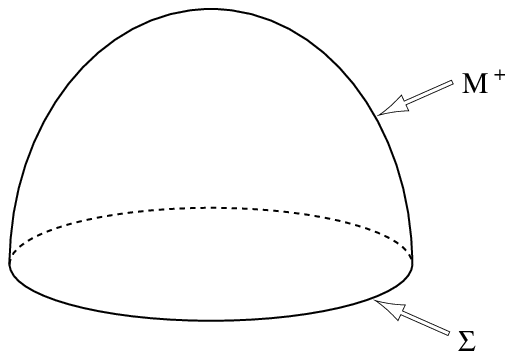}
$${\rm Probability ~of~} h_{ij} ~= ~\Psi^+(h_{ij})\times\Psi^-(h_{ij})$$
\vskip -0.2in
$${\rm where ~}\Psi^+(h_{ij}) = \int_{{\rm metrics ~on}~M^+~{\rm that}
\atop{\rm induce}~h_{ij}~{\rm on}~\Sigma}d[g]~e^{-I}$$
\endinsert

\noindent In this case, the probability for $\Sigma$ to have
the metric $h_{ij}$ can be factorized. It is the product of two wave functions
$\Psi^+$ and $\Psi^-$. These are given by path integrals over
all metrics on $M^+$ and $M^-$ respectively, that induce the given three
metric $h_{ij}$ on $\Sigma$. In most cases, the two wave functions will be
equal
and I will drop the superscripts $+$ and $-$. $\Psi$ is called the wave
function of the universe. If there are matter fields $\phi$, the wave function
will also depend on their values $\phi_0$ on $\Sigma$. But it will not depend
explicitly on time because there is no preferred time coordinate in a closed
universe. The no boundary proposal implies that the wave function of the
universe is given by a path integral over fields on a compact manifold $M^+$
whose only boundary is the surface $\Sigma$. The path integral is taken over
all
metrics and matter fields on $M^+$ that agree with the metric $h_{ij}$ and
matter fields $\phi_0$ on $\Sigma$.

One can describe the position of the surface $\Sigma$ by a function $\tau$ of
three coordinates $x_i$ on $\Sigma$. But the wave function defined by the path
integral can't depend on $\tau$ or on the choice of the coordinates $x_i$.
This
implies that the wave function $\Psi$ has to obey four functional
differential equations. Three of these equations are called the momentum
constraints.

\midinsert
\hsize=4.75in
\indent
\boxit{\smallskip
{\bf Momentum Constraint Equation}
$$\left( {\partial\Psi\over \partial h_{ij}}\right)_{;j} ~ = ~ 0$$
}
\hsize=5.4in
\endinsert

\noindent They express the fact that the wave function should be the same
for different 3 metrics $h_{ij}$ that can be obtained from each other by
transformations of the coordinates $x_i$. The fourth equation is called the
Wheeler-DeWitt equation.

\midinsert
\hsize=4.75in
\indent
\boxit{\smallskip
{\bf Wheeler - DeWitt Equation}
$$\left( G_{ijkl}{\partial^2\over\partial h_{ij}\partial h_{kl}}
-h^{1\over 2} ~{}^3R\right) \Psi~ = ~ 0$$
}
\hsize=5.4in
\endinsert

\noindent It corresponds to the independence of the wave function
on $\tau$. One can think of it as the Schr\"odinger equation for the universe.
But
there is no time derivative term because the wave function does not depend on
time explicitly.

In order to estimate the wave function of the universe, one can use the saddle
point approximation to the path integral as in the case of black holes. One
finds a Euclidean metric $g_0$ on the manifold $M^+$ that satisfies the field
equations and induces the metric $h_{ij}$ on the boundary $\Sigma$. One can
then
expand the action in a power series around the background metric $g_0$.
$$I[g]~=~I[g_0]~+~{1\over 2}\delta g I_2 \delta g ~+~ ...$$

\noindent As
before the term linear in the perturbations vanishes. The quadratic term can
be regarded as giving the contribution of gravitons on the background and the
higher order terms as interactions between the gravitons. These can be ignored
when the radius of curvature of the background is large compared to the Planck
scale. Therefore
$$\Psi~\approx~{1\over ({\rm det}~ I_2 )^{1\over 2}}e^{-I[g_o]}$$

One can see what the wave function is like from a simple example. Consider a
situation in which there are no matter fields but there is a positive
cosmological constant $\Lambda$.

\midinsert
\hskip -0.1in
\epsfbox{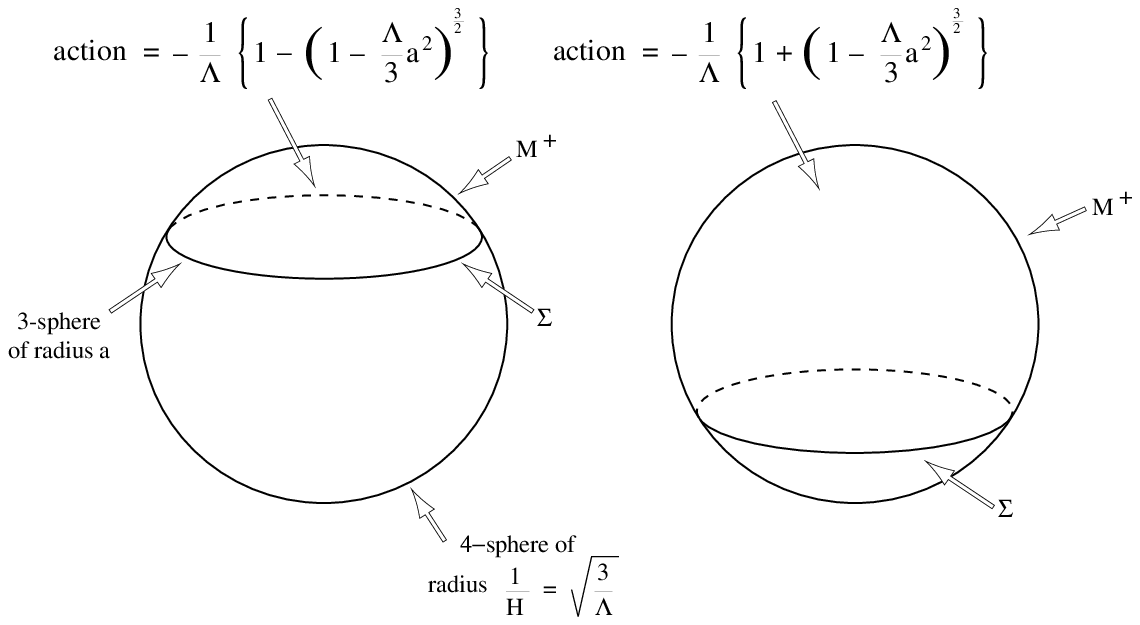}
\endinsert

\noindent  Let us take the surface $\Sigma$ to be a
three sphere and the metric $h_{ij}$ to be the round three sphere metric of
radius $a$. Then the manifold $M^+$ bounded by $\Sigma$ can be taken to be the
four ball. The metric that satisfies the field equations is part of a four
sphere of radius ${1\over H}$ where $H^2 = {\Lambda \over 3}$.
$$I~=~{1\over 16\pi}\int (R-2\Lambda )(-g)^{1\over 2}d^4x~+~
{1\over 8\pi}\int K(\pm h)^{1\over 2} d^3x$$
For a three sphere $\Sigma$ of radius less than ${1 \over H}$ there are two
possible
Euclidean solutions: either $M^+$ can be less than a hemisphere or it can be
more. However there are arguments that show that one should pick the solution
corresponding to less than a hemisphere.

The next figure shows the contribution to the wave function that comes
from the action of the metric $g_0$. When the radius of $\Sigma$ is less than
${1\over
H}$ the wave function increases exponentially like $e^{a^2}$. However,
when $a$ is greater than ${1\over H}$ one can analytically continue the result
for
smaller $a$ and obtain a wave function that oscillates very rapidly.

\midinsert
\hskip 0.7in
\epsfbox{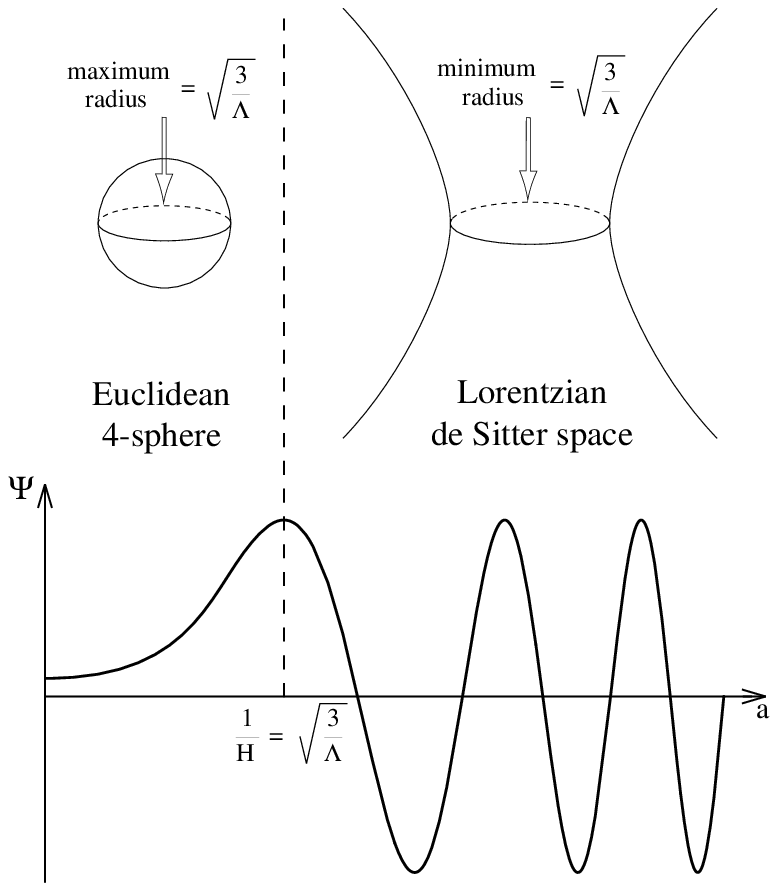}
\endinsert

One can interpret this wave function as follows. The real time solution of the
Einstein equations with a $\Lambda$ term and maximal symmetry is de Sitter
space.
This can be embedded as a hyperboloid in five dimensional Minkowski space.

\midinsert
\hsize=4.75in
\indent
\boxit{\smallskip
{\bf Lorentzian - de Sitter Metric}
$$ds^2=-dt^2~+~{1\over H^2} {\rm cosh} Ht(dr^2+\sin^2r(d\theta^2+\sin^2\theta
d\phi^2))$$
\hskip 1.5in
\epsfbox{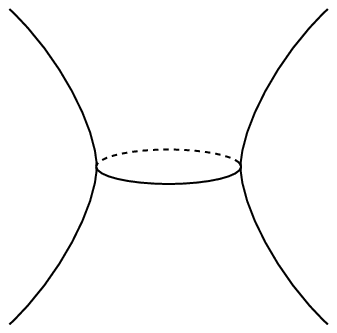}
\smallskip}
\hsize=5.4in
\endinsert

\noindent One
can think of it as a closed universe that shrinks down from infinite size to
a minimum radius and then expands again exponentially. The metric can be
written in the form of a Friedmann universe with scale factor ${\rm cosh} H t$.
Putting $\tau = i t$ converts the ${\rm cosh}$ into $\cos$ giving the Euclidean
metric on a
four sphere of radius ${1\over H}$.

\midinsert
\hsize=4.75in
\indent
\boxit{\smallskip
{\bf Euclidean Metric}
$$ds^2=d\tau^2~+~{1\over H^2} \cos H\tau (dr^2+\sin^2r(d\theta^2+\sin^2\theta
d\phi^2))$$
\hskip 1.7in
\epsfbox{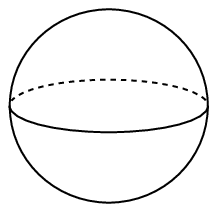}
\smallskip}
\hsize=5.4in
\endinsert

\noindent Thus one gets the idea that a wave function
which varies exponentially with the three metric $h_{ij}$ corresponds to an
imaginary time Euclidean metric. On the other hand, a wave function which
oscillates rapidly corresponds to a real time Lorentzian metric.

Like in the case of the pair creation of black holes, one can describe the
spontaneous creation of an exponentially expanding universe. One joins the
lower
half of the Euclidean four sphere to the upper half of the Lorentzian
hyperboloid.

\midinsert
\hskip 1.3in
\epsfbox{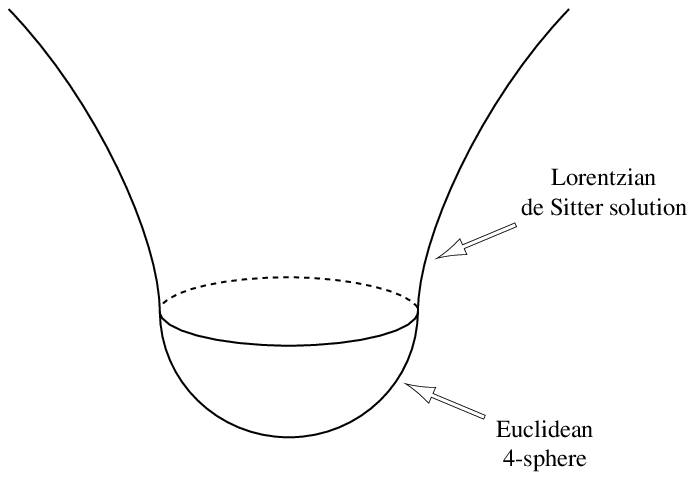}
\endinsert

\noindent Unlike the black hole pair creation, one couldn't say that the de
Sitter universe was created out of field energy in a pre-existing space.
Instead, it would quite literally be created out of nothing: not just out of
the
vacuum but out of absolutely nothing at all because there is nothing outside
the universe. In the Euclidean regime, the de Sitter universe is just a closed
space like the surface of the Earth but with two more dimensions. If the
cosmological constant is small compared to the Planck value, the curvature of
the Euclidean four sphere should be small. This will mean that the saddle point
approximation to the path integral should be good, and that the calculation of
the wave function of the universe won't be affected by our ignorance of what
happens in very high curvatures.

One can also solve the field equations for boundary metrics that aren't exactly
the round three sphere metric. If the radius of the three sphere is less than
${1\over H}$, the solution is a real Euclidean metric. The action will be real
and the wave function will be exponentially damped compared to the round three
sphere of the same volume. If the radius of the three sphere is greater than
this critical radius there will be two complex conjugate solutions and the
wave function will oscillate rapidly with small changes in $h_{ij}$.

Any measurement made in cosmology can be formulated in terms of the wave
function. Thus the no boundary proposal makes cosmology into a science because
one can predict the result of any observation. The case we have just been
considering of no matter fields and just a cosmological constant does not
correspond to the universe we live in. Nevertheless, it is a useful example,
both because it is a simple model that can be solved fairly explicitly and
because, as we shall see, it seems to correspond to the early stages of the
universe.

Although it is not obvious from the wave function, a de Sitter universe has
thermal properties rather like a black hole. One can see this by writing the de
Sitter metric in a static form rather like the Schwarzschild solution.

\midinsert
\hsize=4.75in
\indent
\boxit{\smallskip
{\bf Static form of the de Sitter metric}
$$ds^2=-(1-H^2r^2)dt^2~+~(1-H^2r^2)^{-1} dr^2~+~
r^2(d\theta^2+\sin^2\theta d\phi^2)$$
\hskip 0.65in
\epsfbox{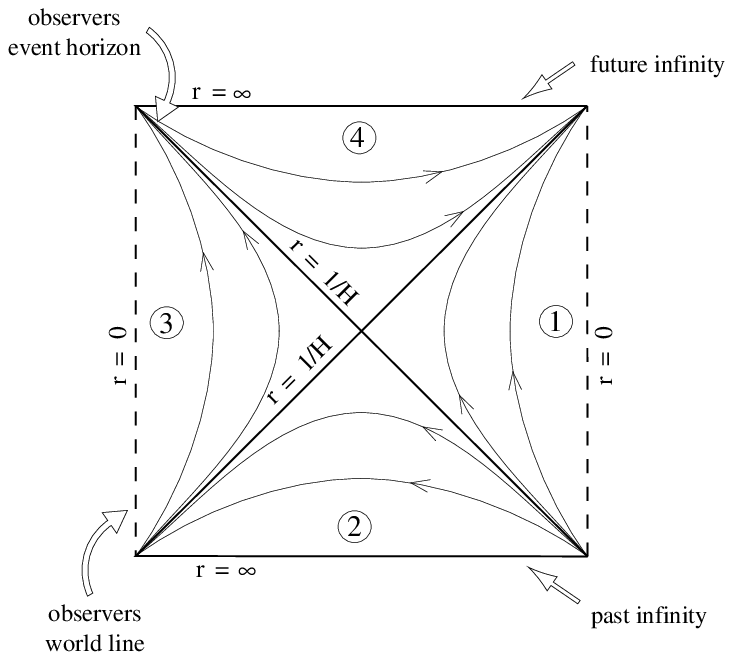}
\smallskip
}
\hsize=5.4in
\endinsert

\noindent There is
an apparent singularity at $r = {1\over H}$. However, as in the Schwarzschild
solution, one can remove it by a coordinate transformation and it corresponds
to an event horizon. This can be seen from the Carter-Penrose diagram which is
a square. The dotted vertical line on the left represents the center of
spherical symmetry where the radius $r$ of the two spheres goes to zero. There
is another center of spherical symmetry represented by the dotted vertical line
on the right. The horizontal lines at the top and bottom represent past and
future infinity which are space like in this case. The diagonal line from top
left to bottom right is the boundary of the past of an observer at the left
hand center of symmetry. Thus it can be called his event horizon. However, an
observer whose world line ends up at a different place on future infinity will
have a different event horizon. Thus event horizons are a personal matter in
de Sitter space.

If one returns to the static form of the de Sitter metric and put $\tau = i t$
one gets a Euclidean metric. There is an apparent singularity on the horizon.
However, by defining a new radial coordinate and identifying $\tau$ with period
${2\pi \over H}$, one gets a regular Euclidean metric which is just the
four sphere. Because the imaginary time coordinate is periodic, de Sitter space
and all quantum fields in it will behave as if they were at a temperature
${H\over 2\pi}$. As we shall see, we can observe the consequences of this
temperature in the fluctuations in the microwave background. One can also apply
arguments similar to the black hole case to the action of the Euclidean-de
Sitter solution. One finds that it has an intrinsic entropy of ${\pi\over
H^2}$,
which is a quarter of the area of the event horizon. Again this
entropy arises for a topological reason: the Euler number of the four sphere is
two. This means that there can not be a global time coordinate on Euclidean-de
Sitter space. One can interpret this cosmological entropy as reflecting an
observers lack of knowledge of the universe beyond his event horizon.

\midinsert
\hsize=4.00in
\indent\indent
\boxit{\smallskip
$${\rm Euclidean~metric ~periodic~ with~ period~} {2\pi\over H}$$
$$\Rightarrow~~~~\left\{~~\matrix{
{\rm Temperature}~ = ~{H\over 2\pi}\cr
{\rm Area ~of ~event ~horizon}~ =~{4\pi\over H^2}\cr
{\rm Entropy}~ = ~{\pi\over H^2}\cr}\right.$$
}
\hsize=5.4in
\endinsert

De Sitter space is not a good model of the universe we live in because it is
empty and it is expanding exponentially. We observe that the universe contains
matter and we deduce from the microwave background and the abundances of light
elements that it must have been much hotter and denser in the past. The
simplest scheme that is consistent with our observations is called the Hot Big
Bang model.

\midinsert
\hskip 0.3in
\epsfbox{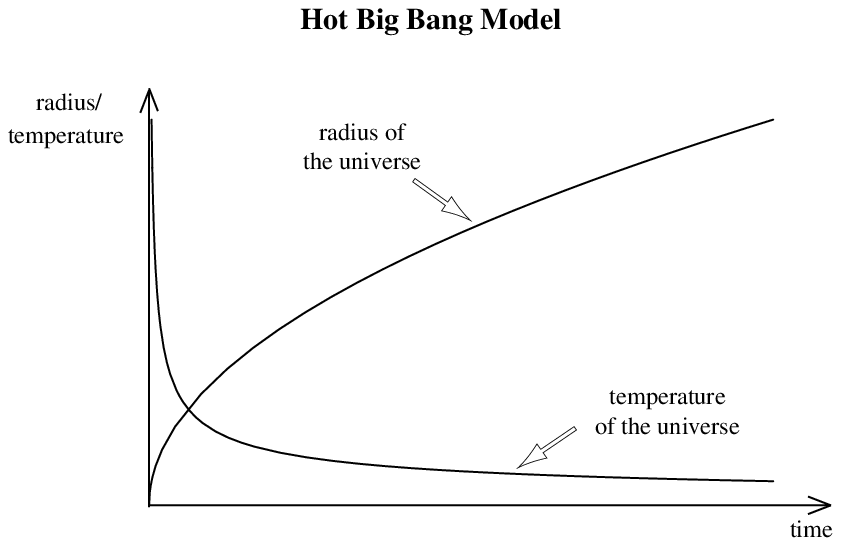}
\endinsert

\noindent In this scenario, the universe starts at a singularity filled with
radiation at an infinite temperature. As it expands, the radiation cools and
its energy density goes down. Eventually the energy density of the radiation
becomes less than the density of non relativistic matter which has dominated
over the expansion by the last factor of a thousand. However we can still
observe the remains of the radiation in a background of microwave radiation at
a temperature of about 3 degrees above absolute zero.

The trouble with the Hot Big Bang model is the trouble with all cosmology
without a theory of initial conditions: it has no predictive power. Because
general relativity would break down at a singularity, anything could come out
of the Big Bang. So why is the universe so homogeneous and isotropic on a large
scale yet with local irregularities like galaxies and stars. And why is the
universe so close to the dividing line between collapsing again and expanding
indefinitely. In order to be as close as we are now the rate of expansion early
on had to be chosen fantastically accurately. If the rate of expansion one
second after the Big Bang had been less by one part in $10^{10}$, the
universe would have collapsed after a few million years. If it had been greater
by one part in $10^{10}$, the universe would have been essentially empty
after a few million years. In neither case would it have lasted long enough for
life to develop. Thus one either has to appeal to the anthropic principle or
find some physical explanation of why the universe is the way it is.

\midinsert
\hsize=4.75in
\indent
\boxit{\smallskip
Hot Big Bang model does not explain why :
\smallskip
\item{1.}The universe is nearly homogeneous and isotropic
but with small perturbations.
\item{2.}The universe is expanding at almost exactly the critical rate
to avoid collapsing again.
\smallskip}
\hsize=5.4in
\endinsert

Some people have claimed that what is called inflation removes the need for a
theory of initial conditions. The idea is that the universe could start out at
the the Big Bang in almost any state. In those parts of the universe in which
conditions were suitable there would be a period of exponential expansion
called inflation. Not only could this increase the size of the region by an
enormous factor like $10^{30}$ or more, it would also leave the region
homogeneous and isotropic and expanding at just the critical rate to avoid
collapsing again. The claim would be that intelligent life would develop only
in regions that inflated. We should not, therefore, be surprised that our
region
is homogeneous and isotropic and is expanding at just the critical rate.

However, inflation alone can not explain the present state of the universe. One
can see this by taking any state for the universe now and running it back in
time. Providing it contains enough matter, the singularity theorems will imply
that there was a singularity in the past. One can choose the initial conditions
of the universe at the Big Bang to be the initial conditions of this model. In
this way, one can show that arbitrary initial conditions at the Big Bang can
lead to any state now. One can't even argue that most initial states lead to a
state like we observe today: the natural measure of both the initial
conditions that do lead to a universe like ours and those that don't is
infinite. One can't therefore claim that one is bigger than the other.

On the other hand, we saw in the case of gravity with a cosmological constant
but no matter fields that the no boundary condition could lead to a universe
that was predictable within the limits of quantum theory. This particular
model did not describe the universe we live in, which is full of matter and
has zero or very small cosmological constant. However one can get a more
realistic model by dropping the cosmological constant and including matter
fields. In particular, one seems to need a scalar field $\phi$ with potential
$V(\phi )$. I shall assume that $V$ has a minimum value of zero at $\phi =0$. A
simple
example would be a massive scalar field $V={1\over 2}m^2\phi^2$.

\midinsert
\hskip 1.2in
\epsfbox{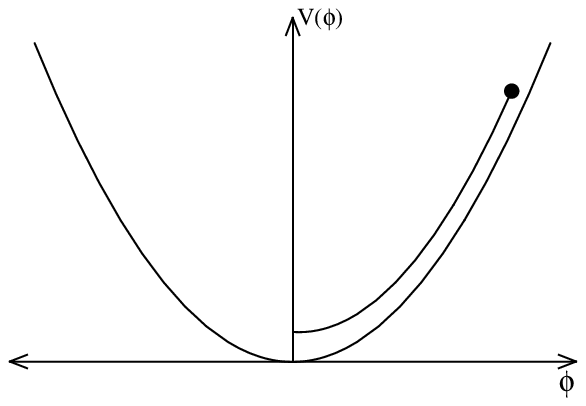}
\endinsert

\midinsert
\hsize=4.75in
\indent
\boxit{\smallskip
{\bf Energy - Momentum Tensor of a Scalar Field}
$$T_{ab}~=~\phi_{,a}\phi_{,b}~-~{1\over 2}g_{ab}\phi_{,c}\phi^{,c}
{}~-~g_{ab}V(\phi )$$
}
\hsize=5.4in
\endinsert

\noindent One
can see from the energy momentum tensor that if the gradient of $\phi$ is small
$V(\phi )$ acts like an effective cosmological constant.

The wave function will now depend on the value $\phi_0$ of $\phi$ on $\Sigma$,
as well
as on the induced metric $h_{ij}$. One can solve the field equations for small
round three sphere metrics and large values of $\phi_0$. The solution with that
boundary is approximately part of a four sphere and a nearly constant $\phi$
field. This is like the de Sitter case with the potential $V(\phi_0)$ playing
the
role of the cosmological constant. Similarly, if the radius $a$ of the three
sphere is a bit bigger than the radius of the Euclidean four sphere there will
be two complex conjugate solutions. These will be like half of the Euclidean
four sphere joined onto a Lorentzian-de Sitter solution with almost constant
$\phi$. Thus the no boundary proposal predicts the spontaneous creation of an
exponentially expanding universe in this model as well as in the de Sitter
case.

One can now consider the evolution of this model. Unlike the de Sitter case, it
will not continue indefinitely with exponential expansion. The scalar field
will
run down the hill of the potential $V$ to the minimum at $\phi =0$. However, if
the
initial value of $\phi$ is larger than the Planck value, the rate of roll down
will be slow compared to the expansion time scale. Thus the universe will
expand
almost exponentially by a large factor. When the scalar field gets down to
order one, it will start to oscillate about $\phi =0$. For most potentials $V$,
the
oscillations will be rapid compared to the expansion time. It is normally
assumed that the energy in these scalar field oscillations will be converted
into pairs of other particles and will heat up the universe. This, however,
depends on an assumption about the arrow of time. I shall come back to this
shortly.

The exponential expansion by a large factor would have left the universe with
almost exactly the critical rate of expansion. Thus the no boundary proposal
can explain why the universe is still so close to the critical rate of
expansion. To see what it predicts for the homogeneity and isotropy of the
universe, one has to consider three metrics $h_{ij}$ which are perturbations of
the round three sphere metric. One can expand these in terms of spherical
harmonics. There are three kinds: scalar harmonics, vector harmonics and tensor
harmonics. The vector harmonics just correspond to changes of the coordinates
$x_i$ on successive three spheres and play no dynamical role. The tensor
harmonics correspond to gravitational waves in the expanding universe, while
the scalar harmonics correspond partly to coordinate freedom and partly to
density perturbations.

\midinsert
\hsize=4.75in
\indent
\boxit{\smallskip
\centerline{Tensor harmonics - Gravitational waves}
\centerline{Vector harmonics - Gauge}
\centerline{Scalar harmonics - Density perturbations}
\smallskip}
\hsize=5.4in
\endinsert

One can write the wave function $\Psi$ as a product of a wave function
$\Psi_0$ for a round three sphere metric of radius $a$ times wave functions for
the
coefficients of the harmonics.
$$\Psi [h_{ij},\phi_0 ]~=~\Psi_0(a,{\bar \phi})
\Psi_a(a_n)\Psi_b(b_n)\Psi_c(c_n)\Psi_d(d_n)$$

\noindent One can then expand the Wheeler-DeWitt equation
for the wave function to all orders in the radius $a$ and the average scalar
field ${\bar \phi}$, but to first order in the perturbations. One gets a series
of
Schr\"odinger equations for the rate of change of the perturbation wave
functions with respect to the time coordinate of the background metric.

\midinsert
\hsize=4.75in
\indent
\boxit{\smallskip
{\bf Schr\"odinger Equations}
$$i {\partial\Psi (d_n)\over\partial t}~=~{1\over 2a^3}\left(
-{\partial^2\over\partial d_n^2}+n^2d_n^2a^4\right)\Psi (d_n)~~~{\rm etc}$$
\smallskip}
\hsize=5.4in
\endinsert

\noindent One can
use the no boundary condition to obtain initial conditions for the perturbation
wave functions. One solves the field equations for a small but slightly
distorted three sphere. This gives the perturbation wave function in the
exponentially expanding period. One then can evolve it using the Schr\"odinger
equation.

The tensor harmonics which correspond to gravitational waves are the
simplest to consider. They don't have any gauge degrees of freedom and they
don't interact directly with the matter perturbations. One can use the no
boundary condition to solve for the initial wave function of the coefficients
$d_n$ of the tensor harmonics in the perturbed metric.

\midinsert
\mathchardef\prp="322F
\hsize=4.75in
\indent
\boxit{\smallskip
{\bf Ground State}
$$\Psi (d_n)~~{\prp}~~e^{-{1\over 2}na^2d_n^2}
 = e^{-{1\over 2}\omega x^2}$$
\vskip -0.2in
$${\rm where}~x~=~a^{3\over 2}d_n~~{\rm and}~~~\omega~=~{n\over a}$$}
\hsize=5.4in
\endinsert

\noindent One finds that it is the
ground state wave function for a harmonic oscillator at the frequency of the
gravitational waves. As the universe expands the frequency will fall. While the
frequency is greater than the expansion rate ${\dot a}/ a$ the Schr\"odinger
equation will allow the wave function to relax adiabatically and the mode will
remain in its ground state. Eventually, however, the frequency will become less
than the expansion rate which is roughly constant during the exponential
expansion. When this happens the Schr\"odinger equation will no longer be able
to change the wave function fast enough that it can remain in the ground state
while the frequency changes. Instead it will freeze in the shape it had when
the frequency fell below the expansion rate.

\midinsert
\hskip 0.4in
\epsfbox{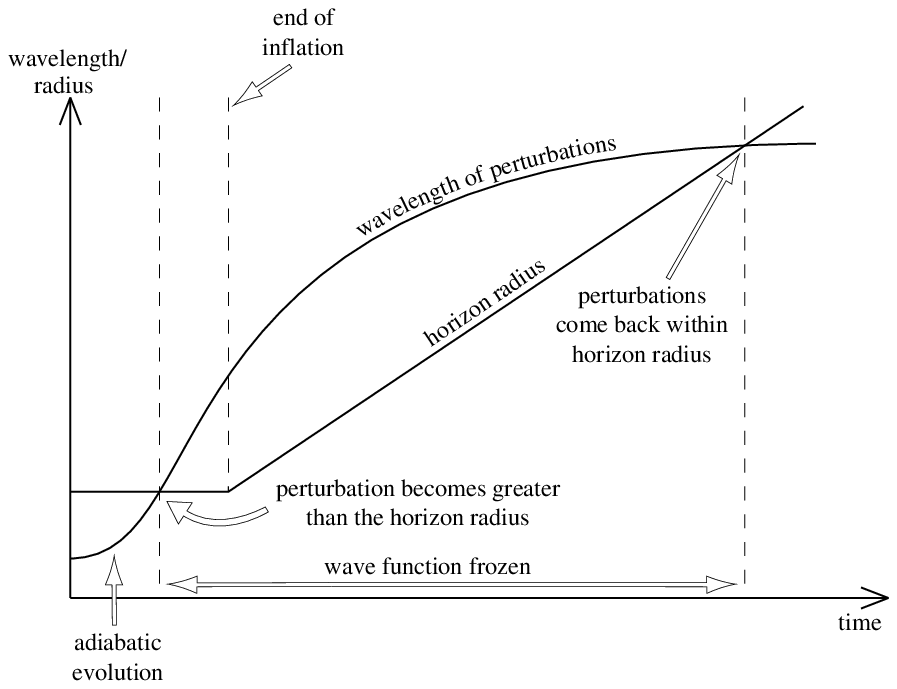}
\endinsert

After the end of the exponential expansion era, the expansion rate will
decrease
faster than the frequency of the mode. This is equivalent to saying that an
observers event horizon, the reciprocal of the expansion rate, increases faster
than the wave length of the mode. Thus the wave length will get longer than the
horizon during the inflation period and will come back within the horizon
later on. When it does, the wave function will still be the same as when the
wave function froze. The frequency, however, will be much lower. The wave
function will therefore correspond to a highly excited state rather than to the
ground state as it did when the wave function froze. These quantum excitations
of the gravitational wave modes will produce angular fluctuations in the
microwave background whose amplitude is the expansion rate (in Planck units) at
the time the wave function froze. Thus the COBE observations of fluctuations
of one part in $10^5$ in the microwave background place an upper limit of
about $10^{-10}$ in Planck units on the energy density when the wave
function froze. This is sufficiently low that the approximations I have used
should be accurate.

However, the gravitational wave tensor harmonics give only an upper limit on
the density at the time of freezing. The reason is that it turns out that the
scalar harmonics give a larger fluctuation in the microwave background. There
are two scalar harmonic degrees of freedom in the three metric $h_{ij}$ and one
in the scalar field. However two of these scalar degrees correspond to
coordinate freedom. Thus there is only one physical scalar degree of freedom
and it corresponds to density perturbations.

The analysis for the scalar perturbations is very similar to that for the
tensor harmonics if one uses one coordinate choice for the period up to the
wave function freezing and another after that. In converting from one
coordinate system to the other, the amplitudes get multiplied by a factor of
the expansion rate divided by the average rate of change of phi. This factor
will depend on the slope of the potential, but will be at least $10$ for
reasonable potentials. This means the fluctuations in the microwave background
that the density perturbations produce will be at least $10$ times bigger than
from the gravitational waves. Thus the upper limit on the energy density at the
time of wave function freezing is only $10^{-12}$ of the Planck
density. This is well within the range of the validity of the approximations I
have been using. Thus it seems we don't need string theory even for the
beginning of the universe.

The spectrum of the fluctuations with angular scale agrees within the accuracy
of the present observations with the prediction that it should be almost scale
free. And the size of the density perturbations is just that required to
explain the formation of galaxies and stars. Thus it seems the no boundary
proposal can explain all the structure of the universe including little
inhomogeneities like ourselves.

One can think of the perturbations in the microwave background as arising from
thermal fluctuations in the scalar field $\phi$. The inflationary period has a
temperature of the expansion rate over $2\pi$ because it is approximately
periodic in imaginary time. Thus, in a sense, we don't need to find a little
primordial black hole: we have already observed an intrinsic gravitational
temperature of about $10^{26}$ degrees, or $10^{-6}$ of the Planck
temperature.

\midinsert
\hsize=4.75in
\indent
\boxit{
$$\eqalign{{{\rm COBE~ predictions ~plus}\atop
{\rm gravitational~ wave~ perturbations}}~&\Rightarrow~
{{\rm upper ~limit ~on ~energy ~density}\atop
10^{-10} ~{\rm Planck~ density}}\cr
{\rm plus~ density ~perturbations}~~~~&\Rightarrow~
{{\rm upper~ limit~ on ~energy ~density}\atop
10^{-12} ~{\rm Planck ~density}}\cr
{{\rm intrinsic~ gravitational}\atop
{\rm temperature~ of ~early~ universe}}~&\approx~
{10^{-6} ~{\rm Planck ~temperature } \atop = ~10^{26}{\rm ~ degrees}}}$$
}
\hsize=5.4in
\endinsert

What about the intrinsic entropy associated with the cosmological event
horizon. Can we observe this. I think we can and that it corresponds to the
fact that objects like galaxies and stars are classical objects even though
they are formed by quantum fluctuations. If one looks at the universe on a
space
like surface $\Sigma$ that spans the whole universe at one time, then it is in
a
single quantum state described by the wave function $\Psi$. However, we
can never see more than half of $\Sigma$ and we are completely ignorant of what
the universe is like beyond our past light cone. This means that in calculating
the probability for observations, we have to sum over all possibilities for the
part of $\Sigma$ we don't observe. The effect of the summation is to change the
part of the universe we observe from a single quantum state to what is called
a mixed state, a statistical ensemble of different possibilities. Such
decoherence, as it is called, is necessary if a system is to behave in a
classical manner rather than a quantum one. People normally try to account for
decoherence by interactions with an external system, such as a heat bath, that
is not measured. In the case of the universe there is no external system, but I
would suggest that the reason we observe classical behavior is that we can see
only part of the universe. One might think that at late times one would be able
to see all the universe and the event horizon would disappear. But this is not
the case. The no boundary proposal implies that the universe is spatially
closed. A closed universe will collapse again before an observer has time to
see all the universe. I have tried to show the entropy of such a universe would
be a quarter of the area of the event horizon at the time of maximum expansion.
However, at the moment, I seem to be getting a factor of ${3\over 16}$ rather
than a ${1\over 4}$. Obviously I'm either on the wrong track or I'm missing
something.

I will end this lecture on a topic on which Roger and I have very different
views, the arrow of time. There is a very clear distinction between the forward
and backward directions of time in our region of the universe. One only has to
watch a film being run backwards to see the difference. Instead of cups falling
off tables and getting broken, they would mend themselves and jump back on the
table. If only real life were like that.

The local laws that physical fields obey are time symmetric, or more precisely,
CPT invariant. Thus the observed difference between the past and the future
must come from the boundary conditions of the universe. Let us take it that the
universe is spatially closed and that it expands to a maximum size and
collapses again. As Roger has emphasized, the universe will be very different
at
the two ends of this history. At what we call the begining of the universe, it
seems to have been very smooth and regular. However, when it collapses again,
we expect it to be very disordered and irregular. Because there are so many
more
disordered configurations than ordered ones, this means that the initial
conditions would have had to be chosen incredibly precisely.

\midinsert
\hskip 0.6in
\epsfbox{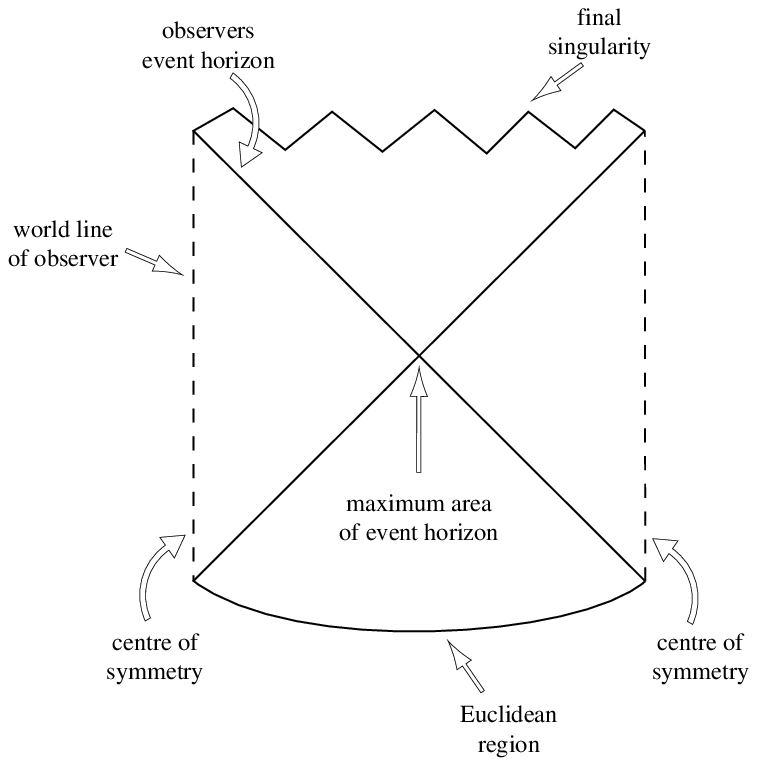}
\endinsert

It seems, therefore, that there must be different boundary conditions at the
two
ends of time. Roger's proposal is that the Weyl tensor should vanish at one end
of time but not the other. The Weyl tensor is that part of the curvature of
spacetime that is not locally determined by the matter through the Einstein
equations. It would have been small in the smooth ordered early stages. But
large in the collapsing universe. Thus this proposal would distinguish the two
ends of time and so might explain the arrow of time.

\midinsert
\hskip 1.4in
\epsfbox{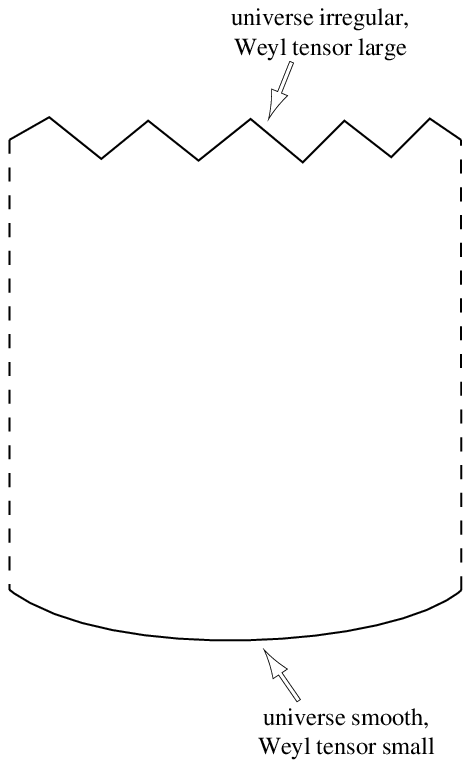}
\endinsert

I think Roger's proposal is Weyl in more than one sense of the word. First, it
is not CPT invariant. Roger sees this as a virtue but I feel one should hang on
to symmetries unless there are compelling reasons to give them up. As I shall
argue, it is not necessary to give up CPT. Second, if the Weyl tensor had been
exactly zero in the early universe it would have been exactly homogeneous and
isotropic and would have remained so for all time. Roger's Weyl hypothesis
could not explain the fluctuations in the background nor the perturbations that
gave rise to galaxies and bodies like ourselves.

\midinsert
\hsize=4.75in
\indent
\boxit{\smallskip
{\bf Objections to Weyl tensor hypothesis}
\smallskip
\item{1.}Not CPT invariant.
\item{2.}Weyl tensor cannot have been exactly zero. Doesn't
explain small fluctuations.
\smallskip}
\hsize=5.4in
\endinsert

Despite all this, I think Roger has put his finger on an important difference
between the two ends of time. But the fact that the Weyl tensor was small at
one
end should not be imposed as an ad hoc boundary condition, but should be
deduced from a more fundamental principle, the no boundary proposal. As we have
seen, this implies that perturbations about half the Euclidean four sphere
joined to half the Lorentzian-de Sitter solution are in their ground state.
That is, they are as small as they can be, consistent with the Uncertainty
Principle. This then would imply Roger's Weyl tensor condition: the Weyl tensor
wouldn't be exactly zero but it would be as near to zero as it could be.

At first I thought that these arguments about perturbations being in their
ground state would apply at both ends of the expansion contraction cycle. The
universe would start smooth and ordered and would get more disordered and
irregular as it expanded. However, I thought it would have to return to a
smooth and
ordered state as it got smaller. This would have implied that the thermodynamic
arrow of time would have to reverse in the contracting phase. Cups would mend
themselves and jump back on the table. People would get younger, not older, as
the universe got smaller again. It is not much good waiting for the universe to
collapse again to return to our youth because it will take too long. But if
the arrow of time reverses when the universe contracts, it might also reverse
inside black holes. However, I wouldn't recommend jumping into a black hole as
a way of prolonging one's life.

I wrote a paper claiming that the arrow of time would reverse when the universe
contracted again. But after that, discussions with Don Page and Raymond
Laflamme convinced me that I had made my greatest mistake, or at least my
greatest mistake in physics: the universe would not return to a smooth state in
the collapse. This would mean that the arrow of time would not reverse. It
would continue pointing in the same direction as in the expansion.

How can the two ends of time be different. Why should perturbations be small at
one end but not the other. The reason is there are two possible complex
solutions of the field equations that match on to a small three sphere
boundary. One is as I have described earlier: it is approximately half the
Euclidean four sphere joined to a small part of the Lorentzian-de Sitter
solution. The other possible solution has the same half Euclidean four sphere
joined to a Lorentzian solution that expands to a very large radius and then
contracts again to the small radius of the given boundary. Obviously, one
solution corresponds to one end of time and the other to the other. The
difference between the two ends comes from the fact that perturbations in the
three metric $h_{ij}$ are heavily damped in the case of the first solution with
only a short Lorentzian period. However the perturbations can be very large
without being significantly damped in the case of the solution that expands and
contracts again. This gives rise to the difference between the two ends of time
that Roger has pointed out. At one end the universe was very smooth and the
Weyl tensor was very small. It could not, however, be exactly zero for that
would have been a violation of the Uncertainty Principle. Instead there would
have been small fluctuations which later grew into galaxies and bodies like
us. By contrast, the universe would have been very irregular and chaotic at the
other end of time with a Weyl tensor that was typically large. This would
explain the observed arrow of time and why cups fall off tables and break
rather than mend themselves and jump back on.

As the arrow of time is not going to reverse, and as I have gone over time, I
better draw my lecture to a close. I have emphasized what I consider the two
most remarkable features that I have learnt in my research on space and time:
first, that gravity curls up spacetime so that it has a begining and an end.
Second, that there is a deep connection between gravity and thermodynamics that
arises because gravity itself determines the topology of the manifold on which
it acts.

The positive curvature of spacetime produced singularities at which classical
general relativity broke down. Cosmic Censorship may shield us from black hole
singularities but we see the Big Bang in full frontal nakedness. Classical
general relativity cannot predict how the universe will begin. However
quantum general relativity, together with the no boundary proposal, predicts a
universe like we observe and even seems to predict the observed spectrum of
fluctuations in the microwave background. However, although the quantum theory
restores the predictability that the classical theory lost, it does not do so
completely. Because we can not see the whole of spacetime on account of black
hole and cosmological event horizons, our observations are described by an
ensemble of quantum states rather than by a single state. This introduces an
extra level of unpredictability but it may also be why the universe appears
classical. This would rescue Schr\"odinger's cat from being half alive and half
dead.

To have removed predictability from physics and then to have put it back again,
but in a reduced sense, is quite a success story. I rest my case.

\end